\documentclass{aa}  
\usepackage{graphicx}
\usepackage{txfonts}
\usepackage{float}
\usepackage[section]{placeins}
\usepackage{enumitem}

\usepackage{graphicx,color}
\usepackage{txfonts}
\usepackage{comment}
\usepackage{amsmath}
\usepackage{xcolor}
\usepackage{hyperref}
\usepackage{caption, multirow}
\DeclareCaptionLabelFormat{bf}{\textbf{#1 #2}}
\usepackage{natbib,twoopt}
\usepackage[hyphenbreaks]{breakurl}

\begin{document}

\title{Optical spectroscopy of high redshift BL Lac objects }

 \author{R. Falomo\inst{1}
          \and
          A. Treves\inst{2,3}
          \and
          S. Paiano\inst{4}
          \and 
          R. Scarpa\inst{5,6}
          }

\institute {INAF - Osservatorio Astronomico di Padova Vicolo dell'Osservatorio 5, I-35122 Padova (PD), Italy\\
\email{renato.falomo@inaf.it}
\and      
INAF - Osservatorio Astronomico di Brera, via Bianchi 46, I-23807, Merate (Lecco), Italy
\and
Università dell'Insubria, via Valleggio 11, 22100, Como, Italy.
\and
INAF - IASF Palermo via Ugo La Malfa 153,I-90146, Palermo Italy        
\and 
Instituto de Astrofisica de Canarias, C/O Via Lactea, s/n E38205, La Laguna (Tenerife), Spain 
\and Universidad de La Laguna, Dpto. Astrofsica, s/n E-38206, La Laguna (Tenerife), Spain 
             }

\date{Received:~; Accepted:~}

\abstract {
BL Lac objects (BLL) are defined by the presence of very weak (typically $<$ 5 Å) or even absent spectral lines. This makes determining their distance particularly challenging, especially at high redshift, where the sources are fainter and the host galaxy contribution in the optical band becomes negligible. Yet measuring their distance is crucial for deriving and modelling their luminosity, notably in the gamma-ray band, where BLLs dominate the extragalactic sky. 
In this work, we re-examine the reported high-redshift (z $>$ 0.6) BLL, many of which are commonly cited in the literature despite appearing questionable. We present new spectra for 52 objects obtained with the 10.4 m GTC. For 16 of them we propose a new redshift, or provide a spectroscopic lower limit, while for 14 sources we confirm previously published values. In 22 cases the spectra remain featureless, even with high S/N observations. These objects are likely to lie at  0.3 $<$  z $<$ 1.4 : the lack of host-galaxy features sets a lower limit to their distance, while the absence of intervening absorption systems argues against substantially higher redshifts. We compare our findings with the previous robustly established cases of BLLs at z $>$ 0.6 that meet our selection criteria.
} 


\maketitle
 
\section{Introduction}  
\label{sec:introduction} 

BL Lac objects (BLL) are active galactic nuclei which are associated with relativistic jets pointing in the direction of the observer and are characterized by very weak or no emission lines in their optical spectra. 
The optical non-thermal continuum is usually well described by a power law and exhibits large and rapid flux variability \citep[see, e.g.,][]{falomo2014}. These sources  dominate the gamma-ray sky: $\sim$50\% of the \textit{Fermi} extragalactic $\gamma$-ray sources are associated with BLL \citep{abdollahi2022, ballet2023}. 
They are supposed to constitute a large fraction, or the majority of the high energy neutrino sources detected by IceCube \citep[see, e.g.,][]{giommi2021}. 

Accurate knowledge of the distance to BLLs is essential in order to model their spectral energy distributions and intrinsic luminosities. The redshift is a key parameter for investigating particle acceleration mechanisms, jet formation, luminosity functions, cosmological evolution, and the contribution of these sources to the extragalactic gamma-ray background. At low redshift, the distance can often be determined through absorption features arising from the host galaxy. At high redshift, however, the situation becomes significantly more challenging. The intrinsic weakness of spectral features, combined with the faintness of the sources and the decreasing contribution of the host galaxy to the observed optical spectrum, makes redshift determination increasingly difficult.

To overcome these limitations, high-quality optical spectroscopy obtained with 10-m class telescopes is required to detect faint spectral features. Nevertheless, even with state-of-the-art instrumentation, emission or absorption lines from the nuclear region remain undetectable in a substantial fraction of high-redshift BLLs.

At high redshift the probability that the BLL light transverses an intervening cloud of gas as in the halo of galaxies becomes relevant. In these cases, the detection of an intervening absorption through recognized absorption lines enables us to establish a spectroscopic lower limit to the redshift. 
Optical spectroscopy of BLL had a great impulse after the launch of the \textit{Fermi} mission in 2008 \citep{atwood2009}, which substantially increased the number of candidate targets, see in particular the extended campaigns by \citet[][]{shaw2012, shaw2013}.
Complementary observations in other bands, as X-rays \citep[see e.g. ][ and references therein]{kerby2021, kaur2023, ulgiati2025} and infrared with WISE \citep[e.g. ][]{massaro2012} allowed the identification of a large number of BLL. 
In preparation of the Cherenkov Telescope Array (CTA), systematic spectroscopic observations of good TeV candidates were made by \citet[][ and references therein]{rajput2025} 
selecting candidates from Monte Carlo simulations of CTAO observations (see details in \cite{goldoni2021} ). 
Very useful data on multifrequency BLL observations up to 2015 are given in the BZCAT catalogue \citet{massaroe2015}. Moreover,  SDSS \footnote{https:/www.sdss4.org/science/data-release-publications} released a substantial number of BLL spectra.

 A number of spectroscopic studies have been carried out to determine spectral properties and the redshift of these sources. See for recent contributions \cite{sbarufatti2005,   sandrinelli2013,pita2014,landoni2018,pena2020,
rajput2025}.

We have contributed to the determination of the redshift of BLL in a number of papers, considering unidentified gamma-ray \textit{Fermi} objects \citep{paiano2017ufo1, paiano2019ufo2,ulgiati2024}, concentrating on extragalactic sources possibly associated with high energy neutrinos \citep{paiano2018txs,paiano2021sin1,paiano2023sin3}, very luminous BLL \citep{paiano2017tev}, and hard \textit{Fermi} sources with counterparts in the TeV region explored by Cherenkov detectors \citep{paiano2020fhl,paiano2021pred}. 
All spectra of BLL studied in the above papers are reported in the ZBLLAC database\footnote{https://web.oapd.inaf.it/zbllac/} \citep{landoni2020}.

In this paper, we focus on high z BLL candidates for which in the literature and/or catalogues there is an indication that the source is at relatively high redshift but, in our opinion, appears dubious based on available data (see details in Sect.~\ref{sec:sample}).  
To investigate these objects, we obtained optical spectra of  55  targets at the Gran Telescopio Canarias. The paper significantly extends a preliminary programme initiated by us \citep{landoni2018, paiano2017high} in which 26 targets were observed.

The paper is organized as follows. In Sect.~\ref{sec:sample} the sample is described together with the adopted selection criteria. Then in Sect.~\ref{sec:obs}, we describe the observations and the spectral analysis.
The results are presented in Sect.~\ref{sec:results} where both the emission or absorption lines and the properties of continuum flux are discussed. For each source, references to previous spectral observations or redshift constrains and individual comments about our spectroscopic data are given in Appendix~\ref{sec:notes}.
Finally, in Sect.~\ref{sec:discuss}, we summarize the results, compare them with previous reliable observations  from the literature, and discuss our findings.
\\

\section{The sample}  
\label{sec:sample} 
We searched for high z ($>$~0.6) BLL selecting sources from the 4LAC Fermi gamma ray catalogue \citep{ajello2020, ajello2022} and the Roma-BZCAT catalogue \citep{massaroe2015} sources at $\delta$$>$~-21$^{\circ}$ as observability condition from the northern hemisphere.
We selected 157 candidates. Then we searched for optical spectra in literature of all these objects, finding at least one published spectrum for 145  targets.
For 57 objects, the spectrum is of enough quality to determine a redshift, or a lower limit ( see Table~\ref{tab:4lac} that reports the redshift and the reference ). These data are discussed in Sect. \ref{sec:discuss}.
For the remaining 88 objects the redshift is uncertain because of low S/N or dubious line identification. We were able to secure optical spectra for 55 targets (see Tab.~\ref{tab:targets}). 
 References for previous spectroscopic observations are given in the last column of Table~\ref{tab:targets}. The reference is given when a spectrum is available in electronic form or a published figure.
For three sources previous good quality spectra were already available and the comparison with the new ones is discussed in the Appendix~\ref{sec:extra}.
 In the following sections we report on the observations and  the results for the sample of 52 targets. 

\section{Observations}  
\label{sec:obs} 

The optical spectra were obtained with the 10.4-m Gran Telescopio Canarias (GTC) at the Roque de los Muchachos Observatory (La Palma), using the OSIRIS spectrograph \citep{cepa2003}. We used the R1000B grism and a 1.2 arcsec slit, providing coverage over 3900–7750~\AA~ with a spectral resolution of R~$\sim$~600. 

Data reduction was carried out using standard IRAF procedures for long-slit spectroscopy \citep{tody1986, tody1993}, following the approach described in \citet{paiano2017tev}. Wavelength calibration was performed using Hg, Ar, Ne, and Xe arc-lamp spectra. The accuracy, based on the scatter of the polynomial fit (pixel versus wavelength), is $\sim$ 0.1 ~\AA~ over the entire spectral range.

To ensure optimal removal of cosmic rays and other artifacts, each target was observed in at least three separate exposures, which also enabled us to identify and discard possible spurious features. The final spectrum was obtained by combining all individual exposures. Atmospheric-extinction correction was applied using the mean extinction coefficients for the La Palma site.

Spectro-photometric standard stars were observed each night to derive the relative flux calibration. The absolute calibration was then refined using the source magnitude measured in the acquisition images. For each acquisition frame, we determined the instrumental magnitude of several stars with catalogued values in SDSS and/or Pan-STARRS, deriving the corresponding calibration constant. In all cases, the resulting accuracy is better than 0.1 mag.

Finally all spectra were dereddened applying the extinction law of \citet{cardelli1989} and assuming the value of Galactic extinction \textit{E(B-V)} derived from the NASA/IPAC Infrared Science Archive\footnote{http://irsa.ipac.caltech.edu/applications/DUST/} \citep{schlafly2011}. 

\section{Results}  
\label{sec:results} 
 For the sample of 52 BLL we report the GTC optical spectrum  in Fig.~\ref{fig:spectra}. 
 The magnitude and S/N of the spectrum can be found in Tab.~\ref{tab:results}. 
Specific details for each source are reported in the Appendix \ref{sec:notes}
, where we also note significant photometric variations relative to previous spectroscopic observations. In particular, when comparing our results with those by \citet{shaw2012,shaw2013}, we follow their indication and assume a spectrophotometric accuracy of 30\%.

\subsection{Redshift and spectral features}

For each spectrum, we searched for emission and absorption features with a (2$\sigma$ ) threshold of minimum equivalent width (EW) that ranges from  0.03  to 2~$\textrm{~\AA}$, depending on the S/N of the spectrum. 
In the cases where a feature is found, we carefully checked that it is present in all individual exposures (see details in Sect.~\ref{sec:obs}).
Close-ups of the main features are shown in Fig.~\ref{fig:zoom}.

For 18 targets we found spectral lines that allowed us to determine the redshift (see Table \ref{tab:results} ). In most cases (15) the redshift was derived from emission lines (see Table~\ref{tab:emlines}), while for 3 targets (5BZBJ0726+3734,  4FGLJ0921.7+2336 and 5BZBJ1132+0034) the absorption features of CaII from the host galaxy were used. 
In 10 of these cases we confirm previous redshift while in 8 cases a new one is proposed (see targets C in Table \ref{tab:results}) 

\begin{figure}
\begin{center}
\includegraphics[width=9cm]{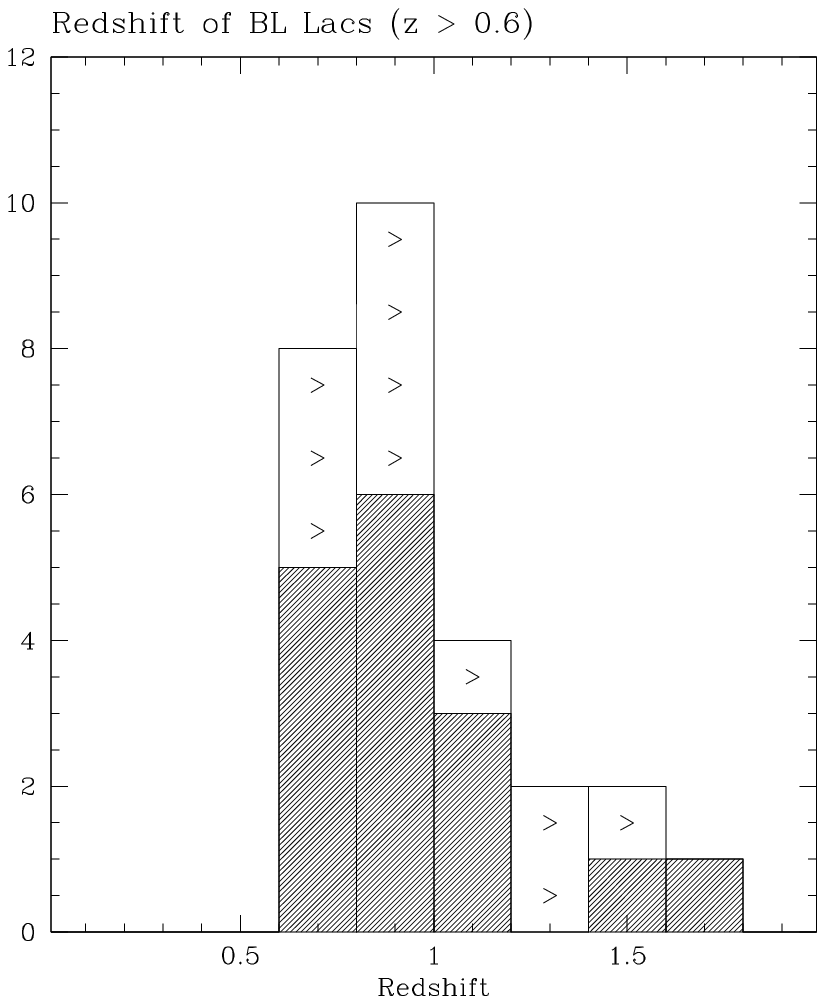}
\caption{
 Distribution of the redshift and the spectroscopic lower limits from  the distribution of intervening systems of BL Lac objects at z $>$ 0.6. See Table \ref{tab:results} for details. } 
\label{fig:zall}  
\end{center}
\end{figure}

For other 12 sources, we detected only intervening absorption lines that set a spectroscopic lower limit to the redshift (see Tab.~\ref{tab:intervening} ).
 In 4 of these cases we confirm previous redshift lower limits while in 8 cases a new limit is proposed (see targets D in Table \ref{tab:results}). 
In Fig.~\ref{fig:zall}, we show the distribution of the redshift and its lower limit for the sample of 27 targets with z $>$ 0.6  (see Table \ref{tab:results}). 
The average redshift is $<z>$~=~0.93~$\pm$~0.30 and the average redshift lower limit is $<z>$~=~1.08~$\pm$~0.35. 

In Tab.~\ref{tab:emlines}, we report measurements of all detected emission lines. 
For 9 objects we detect [OII]~3727 ~\AA. The average luminosity in these sources is $<$~Log$(L[OII])~>$~=~41.5~$\pm$~0.3~erg/s.
This value is very close to the average of the distribution of a collection of $\sim$70 BLL at various redshifts \citep{landoni2020}.
In 7 sources, we detect MgII 2800 broad emission line: in all cases the EW is $<$~10~\AA, but in one case (5BZBJ0407+0742) the EW is $\sim$~17~\AA. 
The luminosity of the line is in the range Log$(L[MgII])~$~=~41.7 to 42.9~erg/s. 

Given the relatively high redshift of these targets, we frequently detected absorption features originating from intervening systems, mainly the MgII~2800 doublet, FeII~2383 and 2600, and in some cases CIV~1550. Table~\ref{tab:intervening} summarizes the properties of all identified intervening absorbers. Overall, at least one such absorption feature is present in 16 of the 52 observed sources.

The most prominent and commonly detected feature is the MgII~2800 doublet and the corresponding redshift range effectively spans from $z \sim 0.4$ to $z \sim 1.7$. On average, MgII absorption is found at $z \sim 0.8 \pm 0.2$, with equivalent widths ranging from 0.3 to 5.7~\AA. The incidence and properties of the MgII intervening systems are consistent with the expectations reported by \citet{Zhu2013} (see also \cite{Landoni2013}).

\subsection{The continuum spectral shape}

\begin{figure}
\includegraphics[width=9cm]{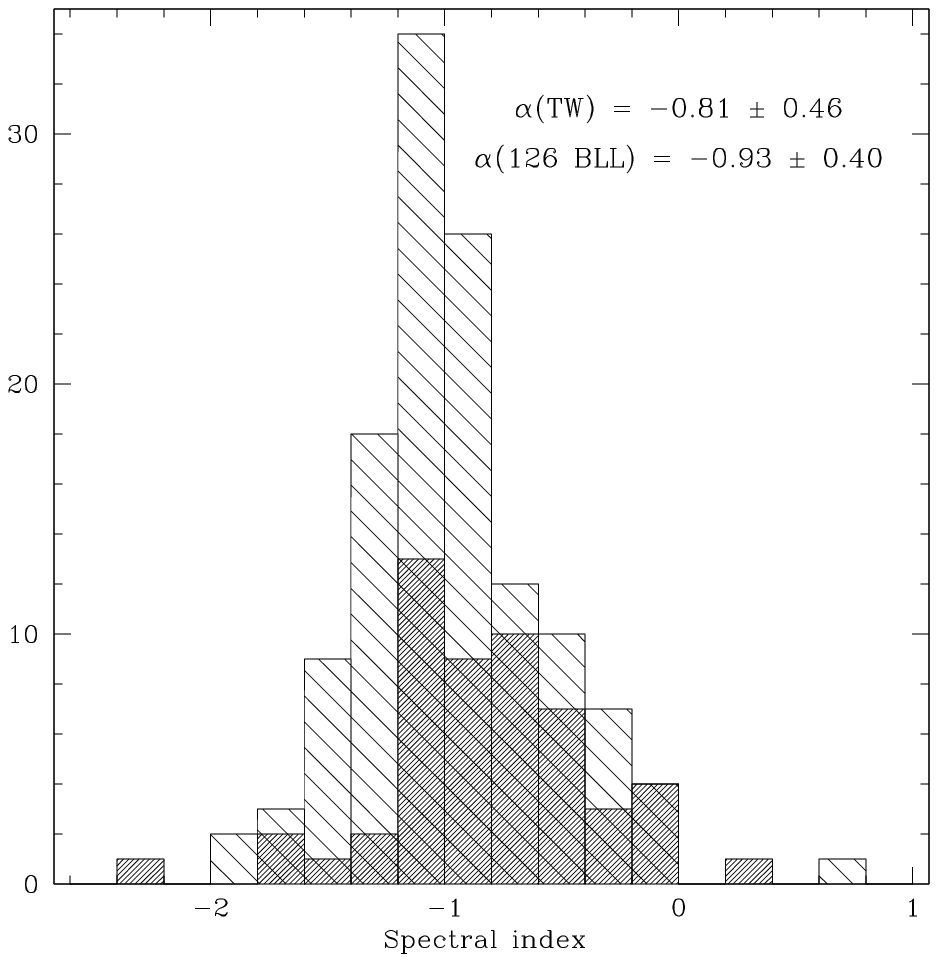}
\caption{
Distribution (filled area) of the spectral index of the optical spectra (4000-7500~\AA)  of the whole dataset (52 BLLs) of this work (TW) compared with the distribution of 126 other BLLs (see papers by Paiano et al referred in Sect. \ref{sec:introduction}). } 
\label{fig:histslp}
\end{figure}

For each source we performed a best fit of the continuum assuming a power law shape after excluding all telluric absorptions and the emission and/or absorption lines in the spectra of targets. We derived the spectral index $\alpha$ from the fit F$_\lambda \propto \lambda^\alpha$ (see Tab.~\ref{tab:results}), binning the spectra in intervals of 100~\AA. 
The spectral index covers a range from -2.3 to 0.23 with an average $<-0.82>~\pm~0.46$ (see Fig.~\ref{fig:histslp}). If we consider only the 22 targets of unknown redshift the average slope becomes $<-1.00>~\pm~0.47$. No significant difference is found from the whole sample. 
In only one case, 4FGLJ1151.5-1347, the global slope of the spectrum is positive and  not well fitted by a power law. 

The distribution of the spectral indexes of this dataset is fully consistent with that of other BLL. In Fig.~\ref{fig:histslp}, we show, for comparison, that of the targets in this work with the spectral index of 126 BLL taken from our previous studies yields the average optical slope of  $\alpha~=~<-0.93>~\pm~0.40$.

\section{Discussion}  
\label{sec:discuss} 

We obtained optical spectra for 52 BLL to probe their high redshift and characterize the optical properties. Only for 14 sources, our observations confirm the redshift (or a spectroscopic lower redshift limit) that was previously reported in the literature (see objects labeled as A in column~6 of Tab.~\ref{tab:results}).
For 8 sources, we disprove the literature redshifts, and propose a new redshift (labeled as C in Tab.~\ref{tab:results}). In 6 of these cases, we found the new redshift is the range 0.7 to 1.5, while in two objects, that were suggested at z$\sim$1, we detected features corresponding to a lower value of z, in the range 0.5 to 0.6. 
In other 8 cases, we do not confirm the previous literature values since no features are detected in our good quality GTC spectra at the proposed redshift.
However, in all these cases, we were able 
to set a spectroscopic lower limit from the detection of intervening absorption lines (see in Tab.~\ref{tab:results} the sources labeled as D). 
These redshift lower limits are in the range 0.66 to 1.80.
 For 22 targets the redshift remains unknown since the spectra are lineless. 

From the present study we are able to derive a firm redshift for 18 sources 
and set lower limits for other 12 objects. 
It is of interest to compare the redshift distribution at $>$ 0.6 of these objects with that of  57 sources for which a reliable redshift or a lower limit  was previously established in the literature (see Sect. \ref{sec:sample} and Tab. \ref{tab:4lac}). 
The average  of 41 literature redshift is $<$0.76 $\pm$ 0.18$>$ ) and the average of 16 redshift lower limit is  $<$0.98 $\pm$ 0.45$>$ ). 
 These values are close to those reported in this paper for 16 sources ($<$0.98 $\pm$ 0.30 $>$ ) and 11 lower limits 
($<$1.08 $\pm$ 0.35 $>$ ).

\begin{figure}
\includegraphics[width=8cm]{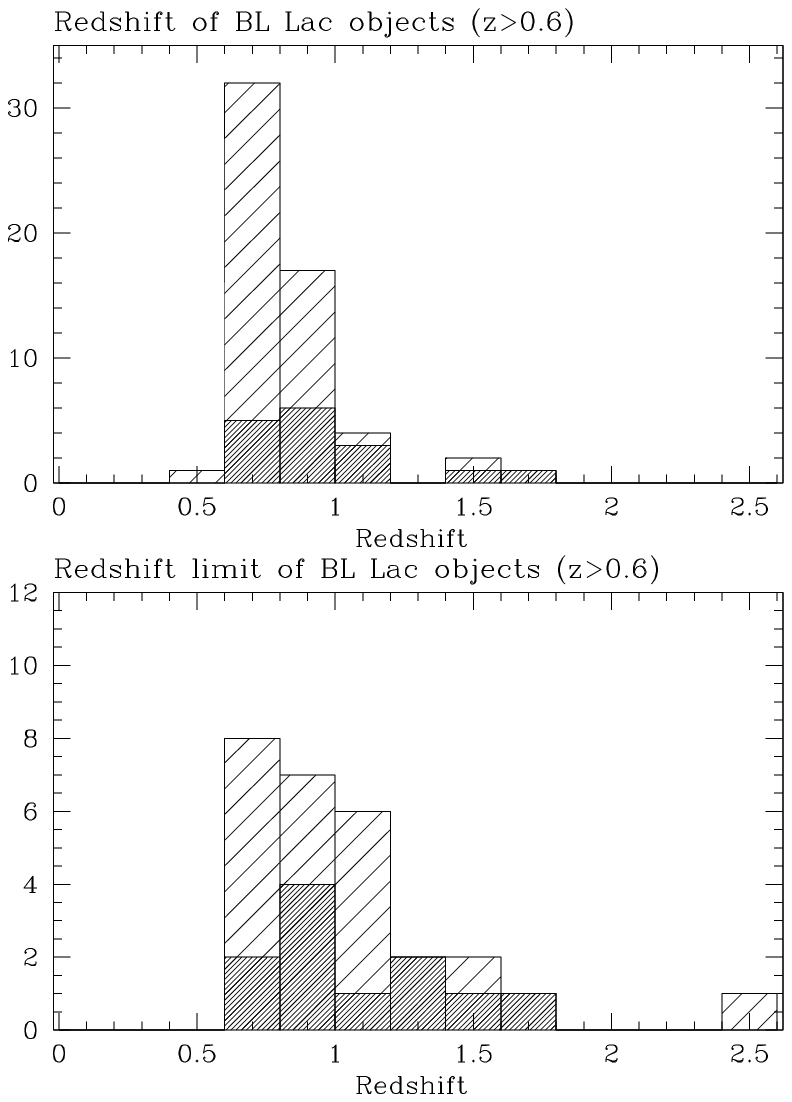}
\caption{ 
{Top:} Redshift distribution of 16  BL Lac objects at z $>$ 0.6 from this work (filled area) and from the whole set of 57 objects including this work and those (41) with known z from Tab.~\ref{tab:4lac} that are  in 4LAC and BZCAT catalogues (shaded area). 
{ Bottom:} Spectroscopic redshift lower limit distribution of 11 objects from this work with z $>$ 0.6 based on intervening absorption systems compared with the lower limit distribution of all 27 sources (objects in this work plus 16  lower limits  in the catalogues  4LAC and BZCAT).}
\label{fig_allz}
\end{figure}

In the top panel of Fig.~\ref{fig_allz}, we show the redshift distribution for the entire sample of 57 objects compared with that from this work.  The concentration of objects at 0.6$<z<$1.0 is likely due to selection effects from flux limited observations. In the bottom panel of Fig.~\ref{fig_allz}, we report the distribution of the redshift lower limits of the entire sample of 27 sources.

From this spectroscopic study of high redshift of BL Lac objects we found 9  sources that are at z $>$ 1. For 5  of them the redshift or its lower limit was previously known. In our previous study of high z BLL candidates (Landoni et al 2018) we found 6 objects at z $>$ 1. Three of them have lower limit set by Ly $\alpha$ absorption at very high z (z = 2.47, 3.14 and 3.36).

Finally, for the remaining 22 objects (labeled as B in Table \ref{tab:results}), we do not confirm the previous alleged values of the redshift and our spectra do not exhibit any emission and/or absorption lines in the spectral range 4000-7800~\AA.
From a careful examination of the spectra reported in the literature we found that in 21 of these cases the proposed redshift  was based on rather uncertain features (see Sect. \ref{sec:notes}). Only in one source (4FGLJ2206.8-0032), the spectrum  by \citet{shaw2013} of October 2010 reports a convincing evidence of an emission line at 5750~\AA, which yields z~=~1.053. In our spectrum (November 2018), which corresponds to a state slightly higher than the \citet{shaw2013} spectrum, this line is undetectable up to EW$\sim$0.3~\AA. Another similar interesting case is 5BZBJ2134--0153 (see details in the Appendix~B3) . This object was observed by us in 2006 (\citet{landoni2018} ) on a flux state that was a factor 5 lower than that of the present observation. In the low state some emission lines are clearly apparent (at z~=~1.285), which are undetected in the spectrum obtained at higher flux  state.

For these 22 sources, characterized by featureless spectra, we derived a minimum redshift based on the non detection of the host galaxy’s spectral signatures. To this end, we first estimated the minimum detectable equivalent width (EW) in each spectrum and then applied the procedure described in \citet{paiano2017tev} to obtain a lower limit on the redshift. 
This method assumes that the observed optical spectrum results from the superposition of a non-thermal power-law component (F$_\lambda \sim \lambda^\alpha$) and the stellar emission of an elliptical host galaxy with absolute magnitude M(R) = –22.9 \citep{sbarufatti2005b}  as derived from HST observations of objects at z $<$ 0.7. Under these assumptions, the detectability of stellar absorption features depends on the nucleus-to-host flux ratio (N/H). Using the observed magnitude of the source and the minimum EW measurable in the spectrum, we can therefore estimate a lower limit to the redshift.

\begin{figure}
\includegraphics[width=9cm]{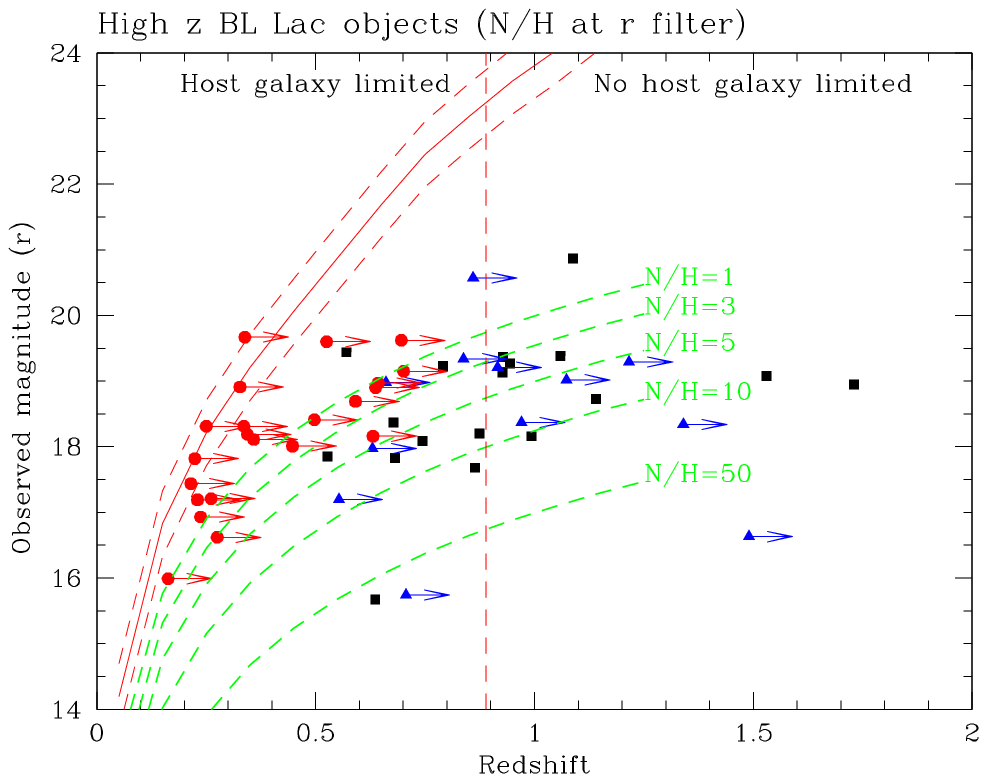}
\caption{The apparent r magnitude of the targets vs the redshift or their lower limits (arrows). Black squares are the objects with known z. The blue triangles with arrow come from spectroscopic lower limits due to intervening absorption systems. The red filled circles with arrows represent the lower limit of the redshift for objects with featureless spectra. These are based on the assumption they are hosted by a typical giant elliptical, that the non thermal nuclear component is described by a power,  and that the CaII absorption features are smaller than the minimum EW of the observed spectra  (see text and \cite{paiano2017tev} for more details). The red dashed vertical line yields the limit of the redshift where CaII lines can be detected in the observe spectral range.
The solid red line represent the magnitude in the r band of a elliptical galaxy of M(R) = -22.9 at various redshift. The two similar red dashed lines encompasses the solid line by 1 magnitude. 
The green dashed lines show the position in the plane z-r of BLL with a standard host galaxy and a non thermal nucleus ($\alpha$ = --1), at different N/H ratio (see text for more details).
}
\label{fig_zm}
\end{figure}

Since the absolute magnitudes of BL Lac host galaxies exhibit a dispersion of roughly one magnitude \citep{sbarufatti2005b}, the computed redshift limits should be regarded as the most probable values. If a host galaxy is 0.5 mag fainter or brighter than assumed, the corresponding redshift limit would increase or decrease by about 0.05–0.1, depending on the redshift. 
 Note that if at z $>$ 0.7 the host were systematically more luminous by $\sim$ 0.5 mag  a further small ($<$ 0.05) different value of redshift could be present.
The resulting lower limits for the sample are shown in Figure \ref{fig_zm} where we also  show the objects with known redshift or with spectroscopic lower limits in the $z$–$r$ plane (redshift versus $r$ magnitude). The $r$ magnitudes were derived from the dereddened spectra. For comparison, we plot the expected apparent $r$ magnitude of the “average” host galaxy (including the $k$-correction) as a function of $z$. We also illustrate the expected total $r$ magnitude (host galaxy plus non-thermal nuclear emission) for several values of N/H in the rest-frame $r$ band.

Most targets with known redshift  lie in the region corresponding to N/H values between 3 and 10. Assuming that the 22 objects with featureless spectra share a similar N/H, these sources would fall within a redshift range of approximately 0.3–1.4 (for an assumed N/H = 5). 
If these objects indeed lie within that redshift interval, one would expect, given high-quality optical spectra, to detect emission lines such as [O II] 3727 ~\AA, [O III] 5007 ~\AA, and Mg II 2800 ~\AA, unless these features are intrinsically very weak. Based on their minimum detectable EW and the assumed redshift range (0.3–1.4), we estimate that the luminosities of these lines would need to be, on average,  lower than $\log(L_{\rm line}) \simeq 40.9$ (with $40.7 < \log(L_{\rm line}) < 41.0$).
This is about an order of magnitude fainter than the average line luminosity reported in Table \ref{tab:emlines}, $\log(L_{\rm line}) = 41.9 \pm 0.5$. Therefore, the possibility to derive their redshift from emission line would require optical spectra of excellent quality in terms of both S/N spectral resolution. 
An alternative approach could rely on the detection of the absorption features from the hosting galaxy. In fact BL Lacs are generally hosted by giant elliptical galaxies, as confirmed in all well-studied cases. Therefore, one of the most effective strategies to determine the redshift of these sources is to obtain high-quality spectroscopy in the red and near-infrared domains, where the host galaxy contributes a larger fraction of the total flux and its main stellar absorption features become more easily detectable. 
It is worth noting that this spectral region is affected by numerous telluric absorption features, which can contaminate the observed spectral lines and therefore require careful correction (see also \cite{pita2014}). In addition, the near-infrared background is significantly higher than in the optical regime, substantially limiting the achievable signal-to-noise ratio of the spectra.
 Finally we comment on another promising technique for constraining the redshift of high z sources relies on the detection of the Lyman break through multicolor photometry (see e.g. \cite{sheng2024} and references therein). However, the accuracy and reliability of the resulting redshift estimates are significantly lower than those obtained from spectroscopic line detections.
 \\

\begin{acknowledgements}
We thank the referee for constructive comments and suggestions. We are grateful to Enrico and Francesco Massaro for clarification on the BZCAT catalogue. 
This paper made use of the Sloan Digital Sky Survey (SDSS) data releases, and of the NASA/IPAC Extragalactic Database (NED).
\end{acknowledgements}

\section*{Data Availability}
The flux-calibrated and de-reddened spectra are available in our online data base ZBLLAC\footnote{\url{http://web.oapd.inaf.it/zbllac/}}.

\bibliographystyle{aa} 
\bibliography{bibliography.bib}

\appendix

\section {Optical spectra}
The GTC optical spectra of high z blazar candidates.

\setcounter{figure}{0}
\begin{figure*}
\centering   
\includegraphics[width=2.0\columnwidth]{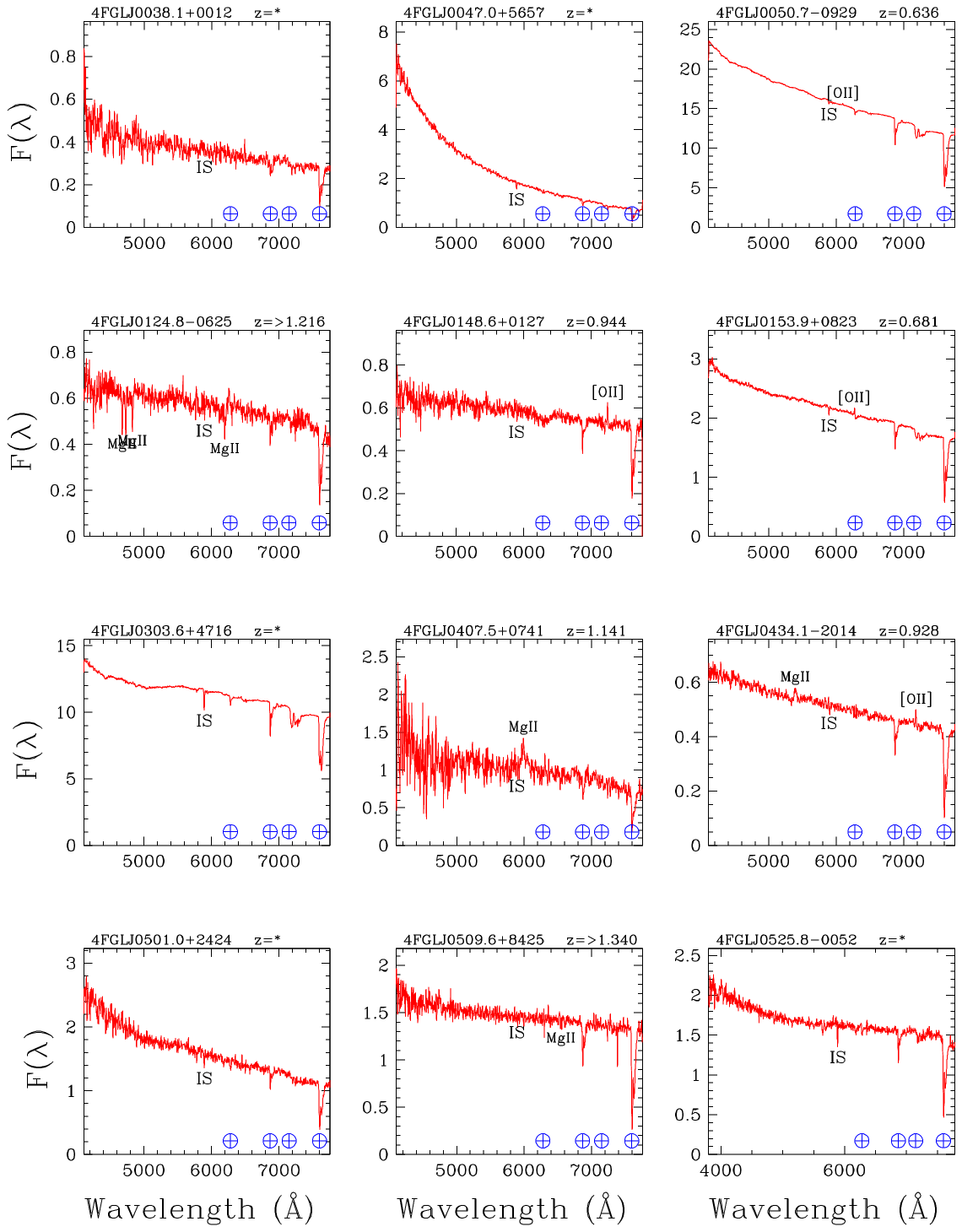} 
\caption{GTC optical spectra of high z blazar candidates (see details in Table \ref{tab:targets} and Table \ref{tab:results} and notes in Appendix \ref{sec:notes}). The redshift of featureless spectra is marked by an asterix. Flux  units :10$^{-16}$ erg cm$^2$ s$^{-1}$ ~\AA$^{-1}$ }
\label{fig:spectra} 
\end{figure*}

\setcounter{figure}{0}
\begin{figure*}
\centering
\includegraphics[width=2.0\columnwidth]{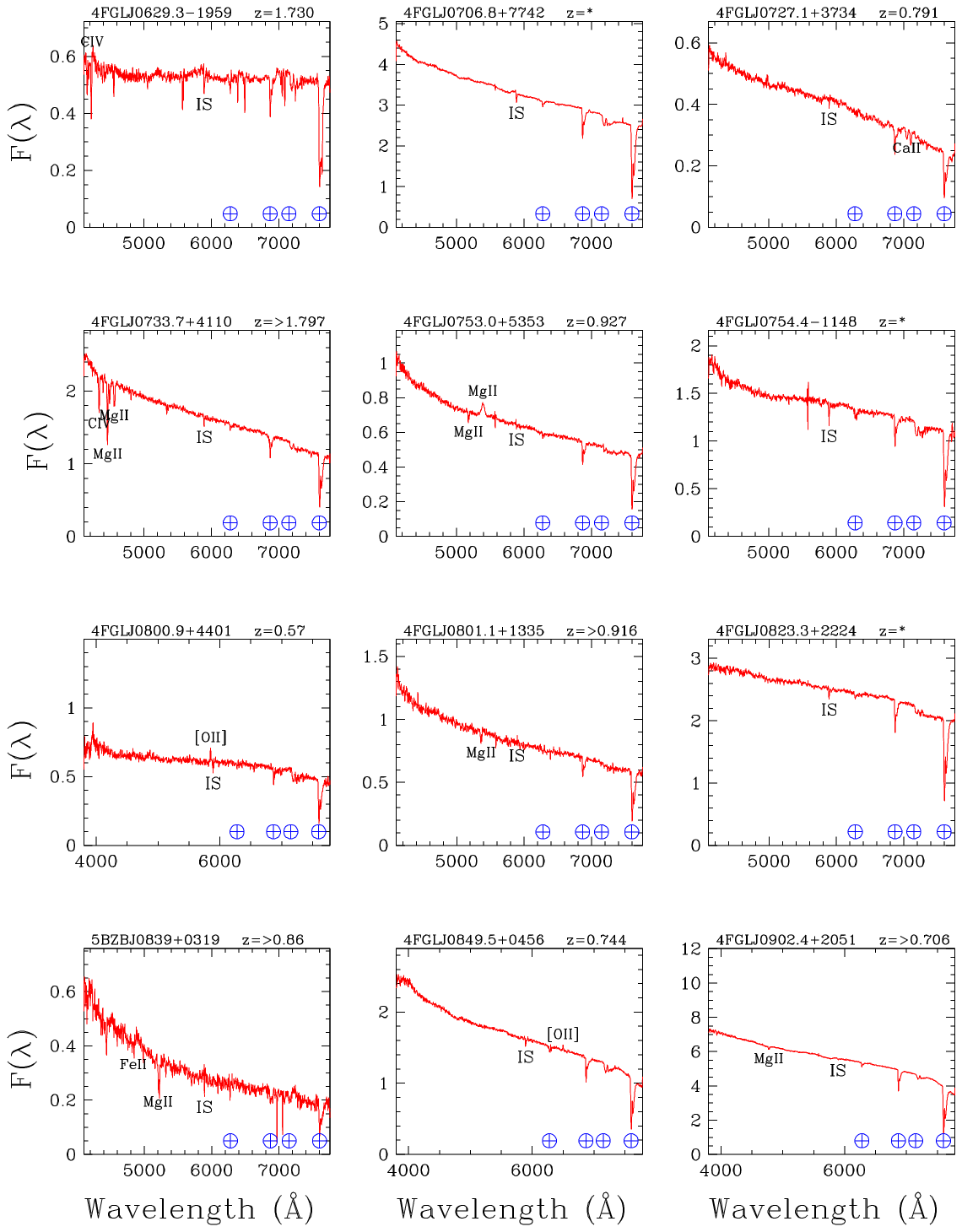} 
\caption{\textit{Continued.}}
\label{fig:spectra}
\end{figure*}

\setcounter{figure}{0}
\begin{figure*}
\centering   
\includegraphics[width=2.0\columnwidth]{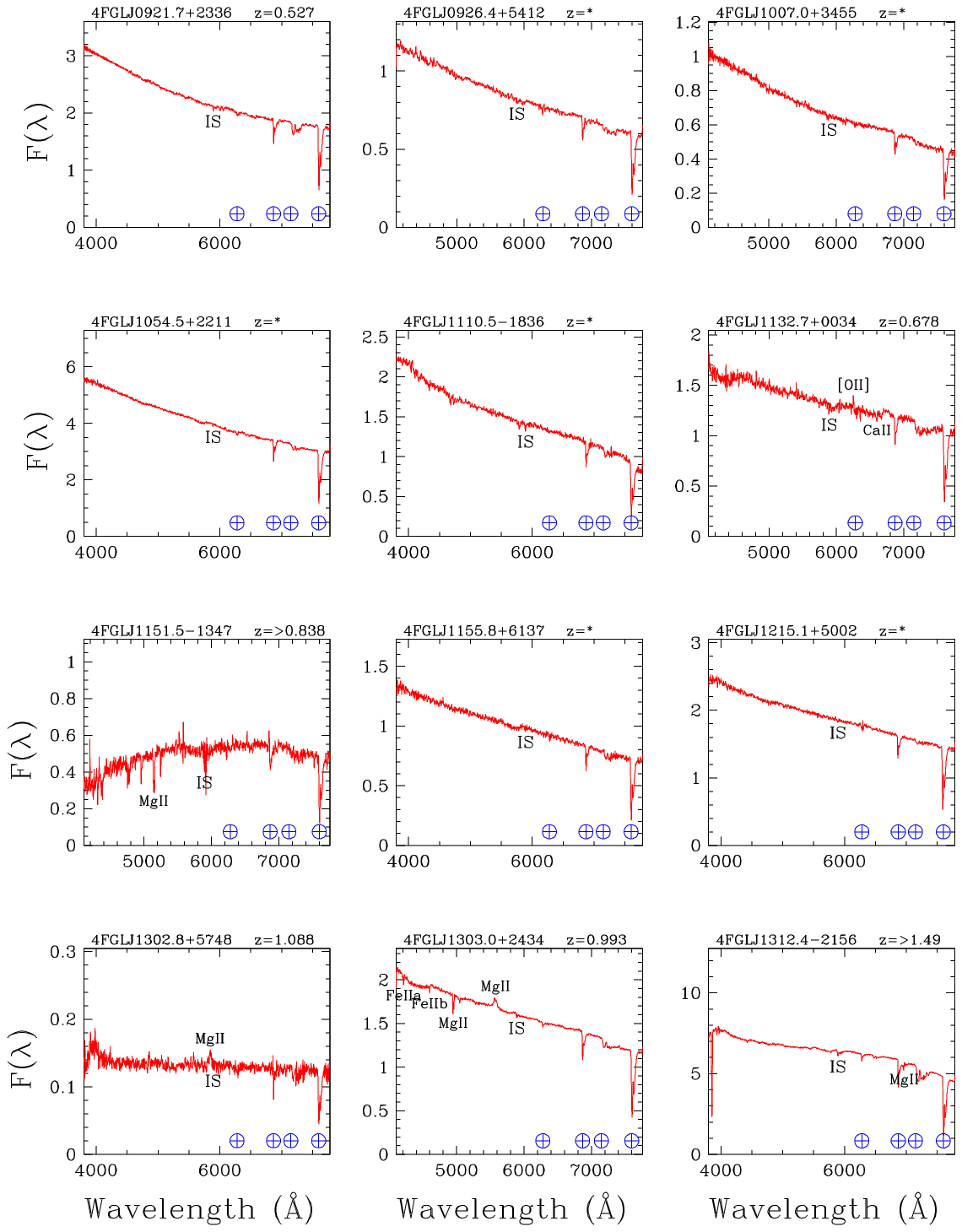} 
\caption{\textit{Continued.}}
\label{fig:spectra}
\end{figure*}

\setcounter{figure}{0}
\begin{figure*} 
\centering   
\includegraphics[width=2.0\columnwidth]{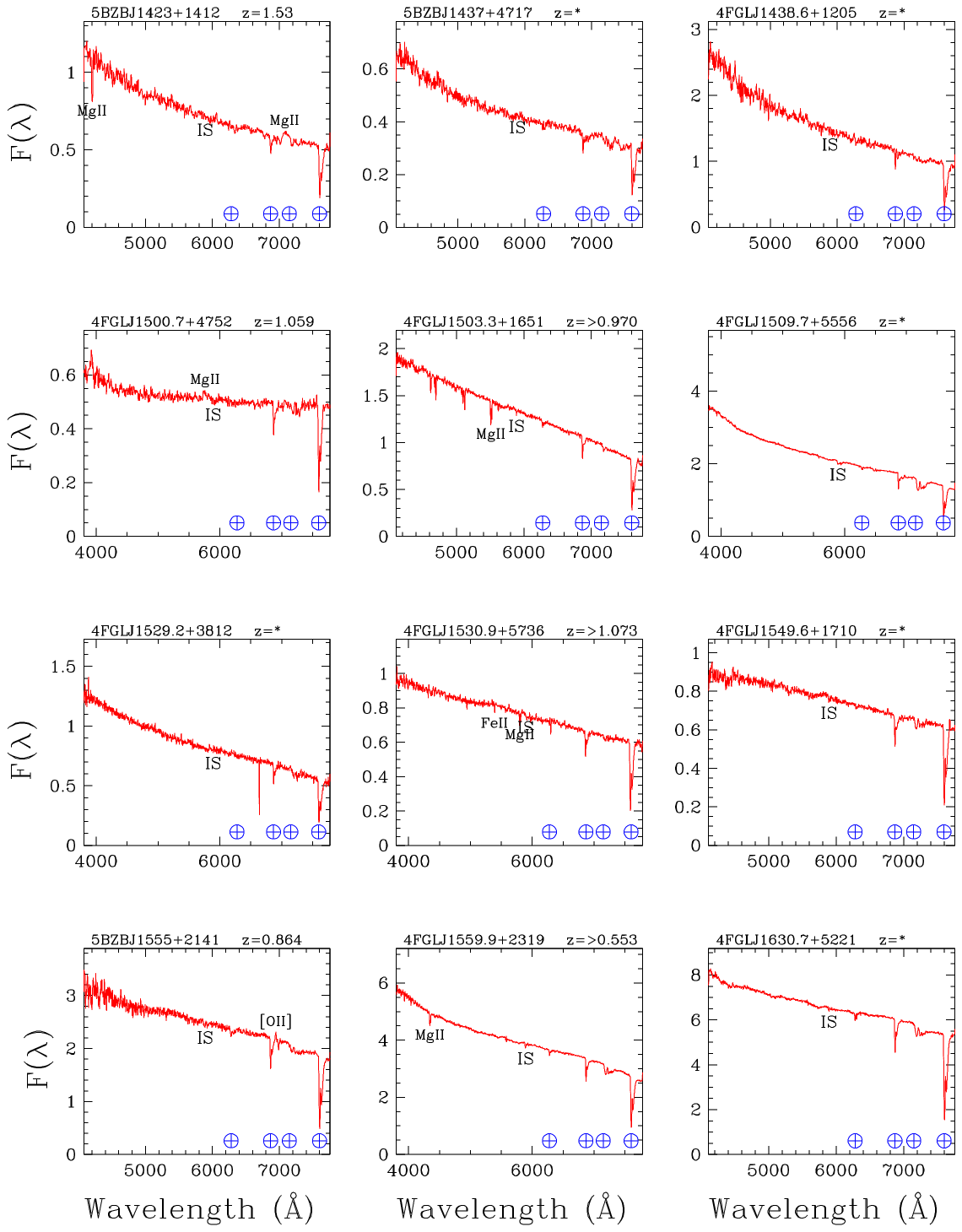} 
\caption{\textit{Continued.}}
\label{fig:spectra}
\end{figure*}

\setcounter{figure}{0}
\begin{figure*} 
\centering   
\includegraphics[width=2.0\columnwidth]{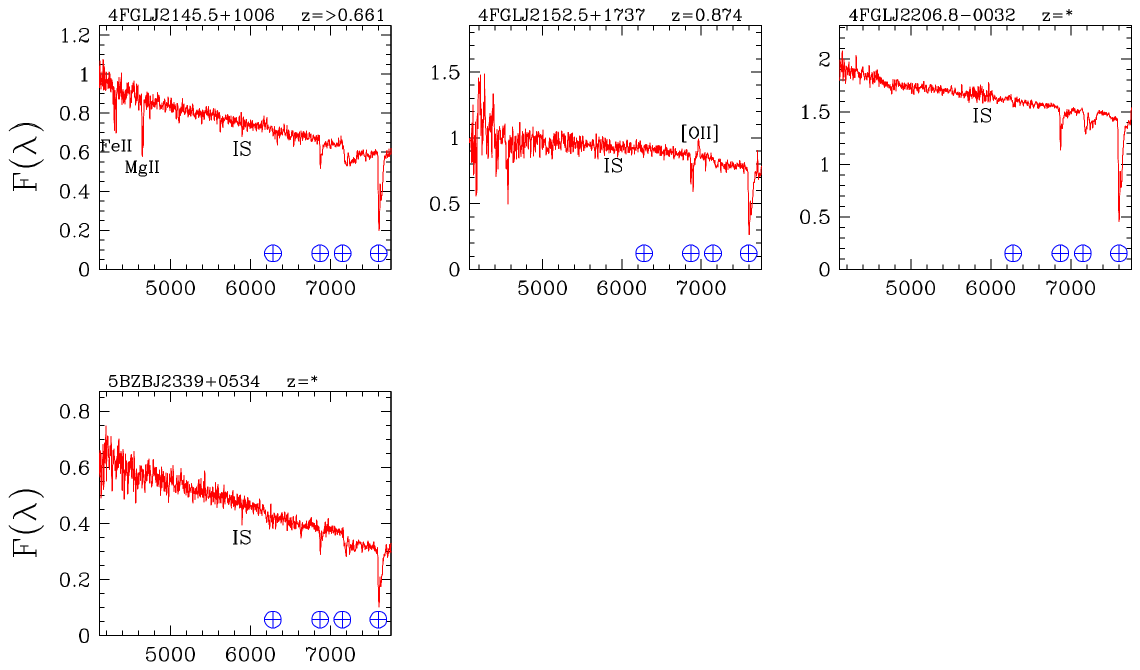} 
\caption{\textit{Continued.}}
\label{fig:spectra}
\end{figure*}

%
\setcounter{figure}{1}
\begin{figure*} \centering   
\includegraphics[width=2.0\columnwidth] 
{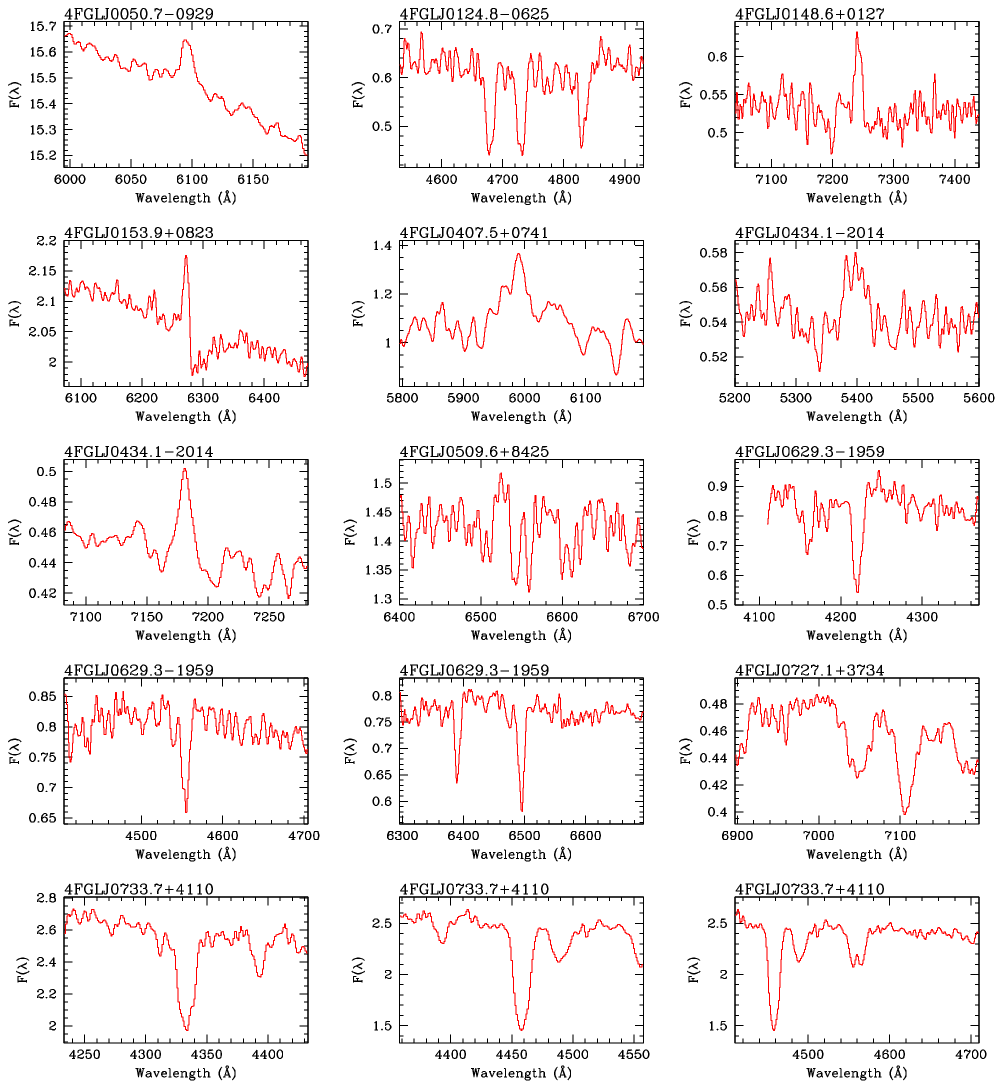}
\caption{Selected spectral regions where relevant emission and absorption lines are detected. See also Tab. \ref{tab:emlines} and Tab. \ref{tab:intervening}  }
\label{fig:zoom} 
\end{figure*}

\setcounter{figure}{1}
\begin{figure*} \centering   
\includegraphics[width=2.0\columnwidth]{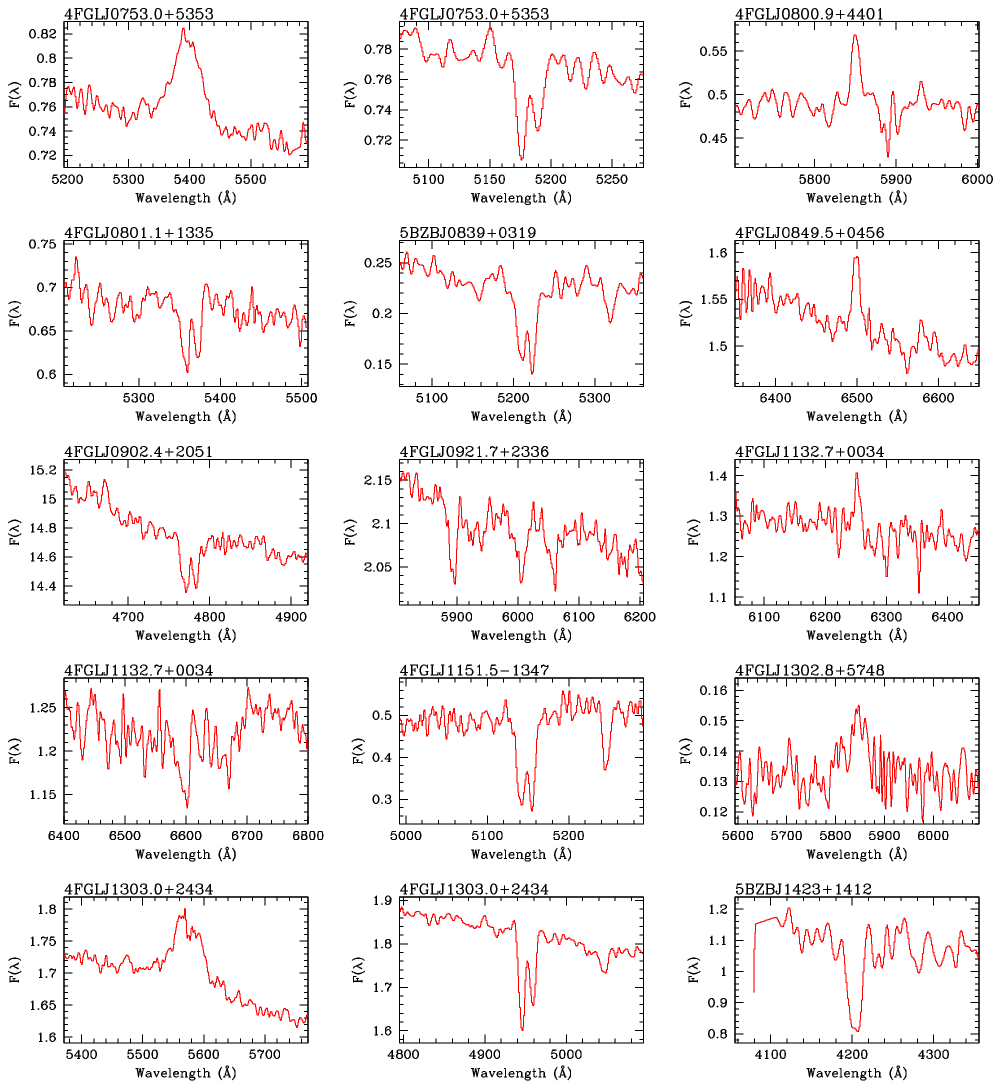}
\caption{Continued}
\end{figure*}

\setcounter{figure}{1}
\begin{figure*} \centering   
\includegraphics[width=2.0\columnwidth]{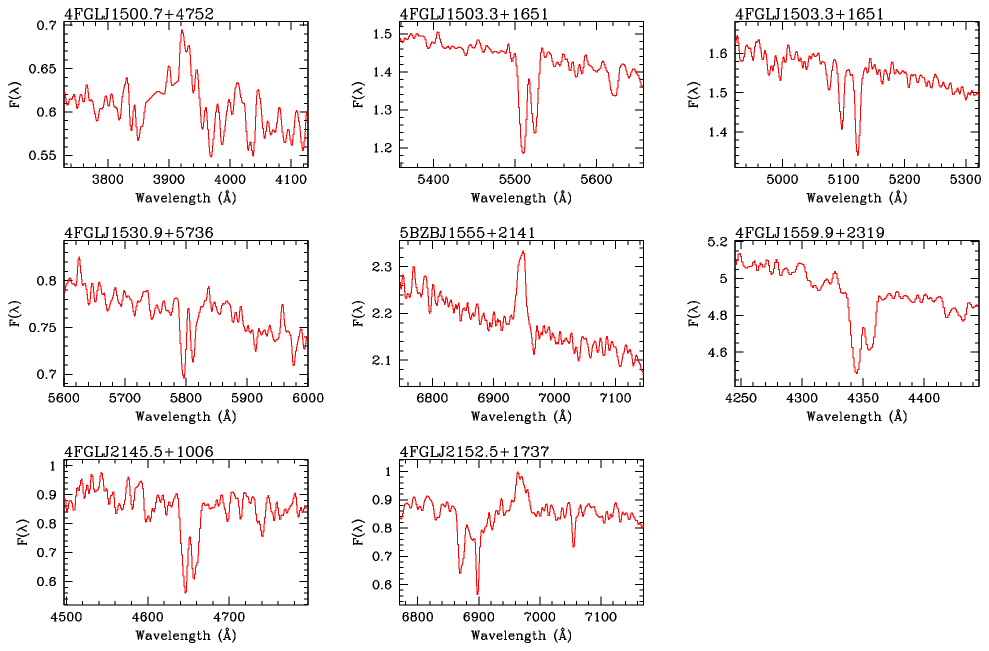}
\caption{Continued}
\end{figure*}

\clearpage
\section{Notes to individual sources} 
\label{sec:notes}

\begin{itemize}

\item[] \textbf{4FGL~J0038.1+0012}: 
In \citet{sandrinelli2013}, a tentative lower limit of the redshift z~$>$~0.708 is proposed based on a possible intervening absorption feature of MgII 2800. SDSS  obtained two spectra  (g=19.7) and proposed z~=~0.739 and z=1.08, but no convincing emission and/or absorption features are clearly visible. Our GTC spectrum was obtained during a faint state  (g=20.2) and is of modest quality (S/N$\sim$15). No emission or absorption lines are found with a limit of EW$\sim$1~\AA.

\item[] \textbf{4FGL~J0047.0+5658}: 
An optical featureless spectrum (g=18.3) was previously obtained by \citet{shaw2013} that estimated a lower limit of the redshift ($>$~0.64) from the absence of host galaxy spectral features. 
Our spectrum was obtained at a lower flux level (g=21.1), and no features are detected at EW~$>$~0.5~\AA.

\allowbreak
\item[] \textbf{4FGL~J0050.7-0929}: 
In the spectrum obtained by \citet{shaw2013}  (g=18.3), two possible faint narrow emission lines are proposed and identified as [OII] and [OIII] at z~=~0.635.
Our better quality spectrum (S/N$\sim$370) clearly shows (see Figure ~\ref{fig:zoom} ) an emission line at 6095~\AA ~(EW~=~0.13~\AA) that confirms the feature proposed in literature. If identified as [OII] 3727 ~\AA~ the redshift is z~=~0.636. Note that the flux level in our spectrum is a factor $\sim$ 7 higher that that obtained by \citet{shaw2013}. 

\item[] \textbf{4FGL~J0124.8-0625}: 
In the spectrum obtained by \citet{shaw2013} many intervening absorption features are visible and identified as CIV~1550 and MgII~2800. A redshift z~=~2.117 is proposed based also on the possible identification of Ly$\alpha$ forest. No clear emission features at the above redshift are visible (and identified) in their spectrum. 
In our GTC spectrum, we clearly detect three narrow absorption features at 4678~\AA, 4730~\AA, and 4830~\AA~ (EW~$\sim$~3.5~\AA ). 
The most probable identification of these three features is in terms of intervening MgII~2800 systems at z~=~0.672, 0.690 and 0.726 (see Fig.~\ref{fig:0125}). These absorption lines are also visible in the spectrum obtained by \citet{shaw2013}, but identified as CIV 1550. 
In our spectrum, another weaker absorption doublet is present at 6197~\AA. If identified as MgII~2800~\AA~, the  intervening system would be  z~=~1.216. Thus this source is at z $>$ 1.216.
\\
\begin{figure}
\includegraphics[width=8.5cm]{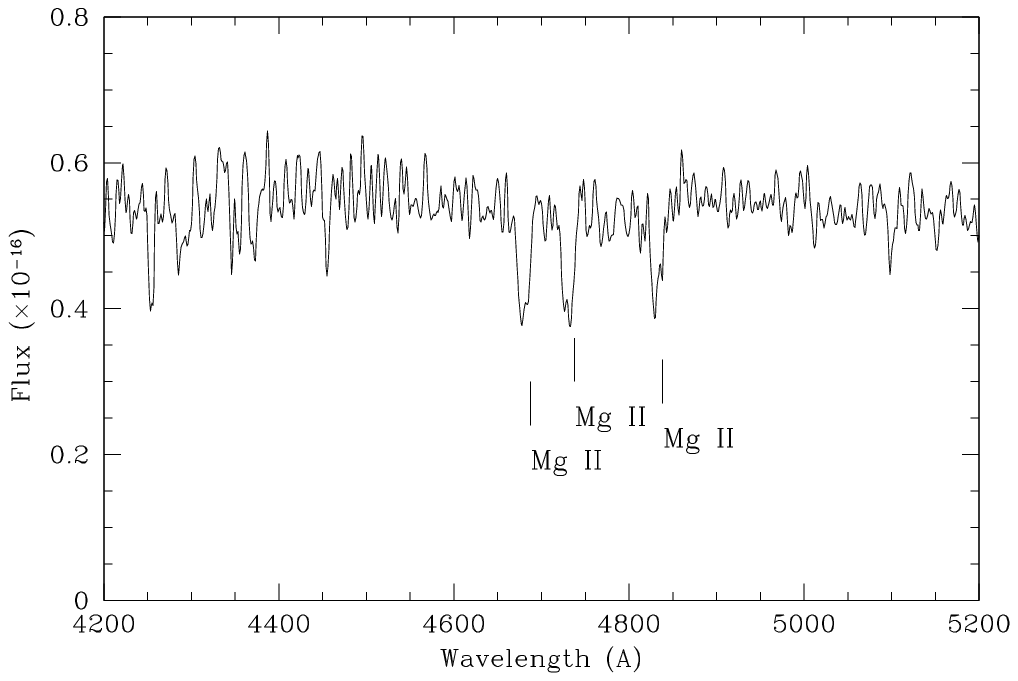}
\caption{The blue region of the GTC optical spectrum of 4FGL~J0124.8-0625. Three absorption features (EW $\sim$ 3.5 ~\AA ) are clearly detected. The most probable identification of these features is intervening MgII 2800 at z=0.67-0.72 }
\label{fig:0125}
\end{figure}

\item[] \textbf{4FGL~J0148.6+0127}:
The spectrum obtained by \citet{shaw2013} is well described by a power law shape. From two spikes identified as [OII] and [OIII] they proposed z~=~0.94.
In our spectrum (S/N$\sim$30), we confirm the detection of a narrow emission line at 7240 (EW$\sim$~3~\AA ). In addition, we also detect a possible weak emission at 5445~\AA. 
These emission lines are identified as [OII]~3727 and MgII~2800, respectively, at z~=~0.944.
\\

\item[] \textbf{4FGL~J0153.9+0823}: 
In the \citet{shaw2013} spectrum, the authors noted two emission lines at $\sim$8416~\AA~ identified as [OIII] at z~=~0.681. These lines are out of observed range of our GTC spectrum, but we clearly detect a narrow emission at 6272~\AA~ (EW~=~0.6~\AA~) that corresponds at the [OII] 3727~\AA~ emission line at the same redshift proposed in literature.  Thus this redshift appears well established from two narrow emission lines. 
\\

\item[] \textbf{4FGL~J0303.6+4716}:
This source is heavily reddened with the 
\textit{E(B-V)}~=~0.22. 
In a spectrum extending to shorter wavelength down to $\sim$3500, \citet{shaw2013} reports an absorption doublet at $\sim$3840~\AA~ that could be identified as MgII 2800. This sets a redshift lower limit of the source at z~$>$~0.37. 
In our good S/N$\sim$250 optical spectrum, we do not detect clear emission or absorption lines down to EW~$>$~0.2~\AA~ in the observed spectral range.
Note that the intervening feature reported by \citet{shaw2013} is outside our observed range. 
\\

\item[] \textbf{4FGLJ0407.5+0741}: 
This source is heavily reddened with the 
\textit{E(B-V)}~=~0.23.
\citet{sowards2003} obtained two optical spectra of the source in two different states that appeared very different in shape and weak features. 
They consider the value z~=~1.133 from the lower state spectrum to be the most reliable.
Our very modest quality (S/N$\sim$15) spectrum of the target (g~=~20.8) shows only one broad emission line (EW~=~22~\AA~) at 5990~\AA~ that, if identified as MgII 2800, yields z~=~1.141.
\\

\item[] \textbf{4FGL~J0434.1-2014}: 
The spectrum obtained by \citet{shaw2013} is well described by a power law shape. Weak emission lines were detected and identified as MgII 2800 and [OII] 3727 at z = 0.928.
In our GTC spectrum, we detect and confirm these two emission lines at 5400~\AA~ and 7182~\AA~ with EW~=~2.5~\AA~ and 1.7~\AA~ respectively, consistent with the a redshift of z~=~0.928.
\\

\item[] \textbf{4FGL~J0501.0+2424}: 
This BLL is reported in the 4FGL and 3FHL catalogue. The source has an unknown redshift and it was studied in the optical-UV band by \citet{sheng2024}. 
This source is heavily reddened ( E(B-V)~=~0.44 ). 
No previous optical spectra of the source are found. 
We obtained an optical spectrum at the GTC that appears featureless with a limit of equivalent width of lines EW~$>$~2~\AA~.
\\

\item[] \textbf{4FGLJ0509.6+8425}: 
The source, also known as S5~0454+84, is listed in the BZCAT catalogue with a reported redshift of z~$>$~1.34, based on the detection of an intervening MgII~2800 absorption system at $\sim$6550~\AA~ \citep{stocke1997}. 
Our GTC optical spectrum (g~=~19.1) is well described by a power law continuum with spectral index $\alpha$$\sim$-0.3. No emission line (EW~$>$~3~\AA~ ) are detected. 
We confirm the presence of a very  weak MgII~2800 absorption doublet at z~=~1.340. \\

\item[] \textbf{4FGL~J0525.8-0052}: 
In the 4LAC catalogue, the source is reported to have a redshift of z~=~1.2, although no reference of spectroscopy is available.
Our GTC optical (r~=~18.5) spectrum is well fitted by a power law shape with spectral index $\alpha$$\sim$-0.6, and no emission and/or absorption lines (EW~$>$~3~\AA~) are detected.\\

\item[] \textbf{4FGL~J0629.3-1959 }
The spectrum obtained by \citet{shaw2013} exhibits many intervening absorption line systems. 
Based on a weak emission feature at $\sim$4225~\AA, identified as CIV~1550~\AA~, they proposed a redshift of z~=~1.724. In our GTC optical spectrum, we confirm the emission feature at 4235~\AA~ (EW~$>$~5~\AA~). This emission is blended with a strong narrow absorption line at 4220~\AA~. 
In addition, we detect another weak emission line at 5202~\AA~ (EW~$>$~2.5~\AA~). These two weak lines are identified as CIV~1550 and CIII]~1909, respectively, yielding a redshift of z~=~1.730 for the source. Furthermore, several intervening absorption system are detected at 4162, 4220, 4555~\AA~, ad well as at 6390, 6495, 7050, and 7088~\AA~. If the absorption features observed at $\lambda$$>$~4500~\AA~ are interpreted as MgII~2800 systems, they correspond to intervening absorbers at redshift between 0.63 and 1.53, consistent with the emission redshift z~=~1.730.
\\
\item[] \textbf{4FGL~J0706.8+7742}: 
In the  power-law continuum obtained by \citet{shaw2013}, no features were detected, and a lower limit of z~$>$~0.39 was proposed. 
Our high quality (S/N$\sim$300) GTC spectrum is also dominated by a power law continuum ($\alpha$$\sim$-1) and appears featureless up to 7700~\AA~. 
\\

\item[] \textbf{4FGL~J0727.1+3734}:
The automatic SDSS analysis of the optical spectrum suggests a redshift of z~=~0.872, based on a single broad emission line at $\sim$7240~\AA, attributed to [NeIII]~3869~\AA~. 
\cite{landoni2018} proposed a redshift z=0.791 on the basis of an absorption doublet at lambda  7043-7107 ~\AA~ identified with CaII 3934-3968. With the present spectrum we confirm the CaII doublet (EW$\sim$3~\AA). Moreover we detect an additional absorption feature at 7348 corresponding to H$_\delta$ at  the same redshift. No emission lines are apparent.
\\

\item[] \textbf{4FGL~J0733.7+4110}: 
The SDSS spectrum tentatively suggests a redshift of z~=~0.19. However, no clear emission features are visible to justify this value.
Our optical spectrum clearly shows a number of absorption features in the spectral range 4200 to 4900~\AA~. The most prominent lines are at 4333~\AA~ (EW~=~3.2~\AA~), 4458~\AA~ (EW~5.7~\AA~) and 4560~\AA~ (EW~3.1~\AA~). 
They are also visible in the SDSS spectrum, which, although noisier than the GTC spectrum,  allows  us to better resolve the features that are doublets in all cases (see Fig.~\ref{fig_0733}).
 We identify these doublets as intervening absorption systems due to one CIV~1550 and two MgII~2800 at redshifts z~=~1.797, 0.594 and 0.629. 
Three additional weaker absorption lines are detected at 4393, 4492 and 4808 ~\AA~, but their identification remains unknown. 
Based on these  features, we can set a lower limit on the source redshift of z~$>$~1.797. Note that at this high z the source becomes very luminous (M$_g < $ --27.2) 

\begin{figure}[H]
\includegraphics[width=9.0cm]{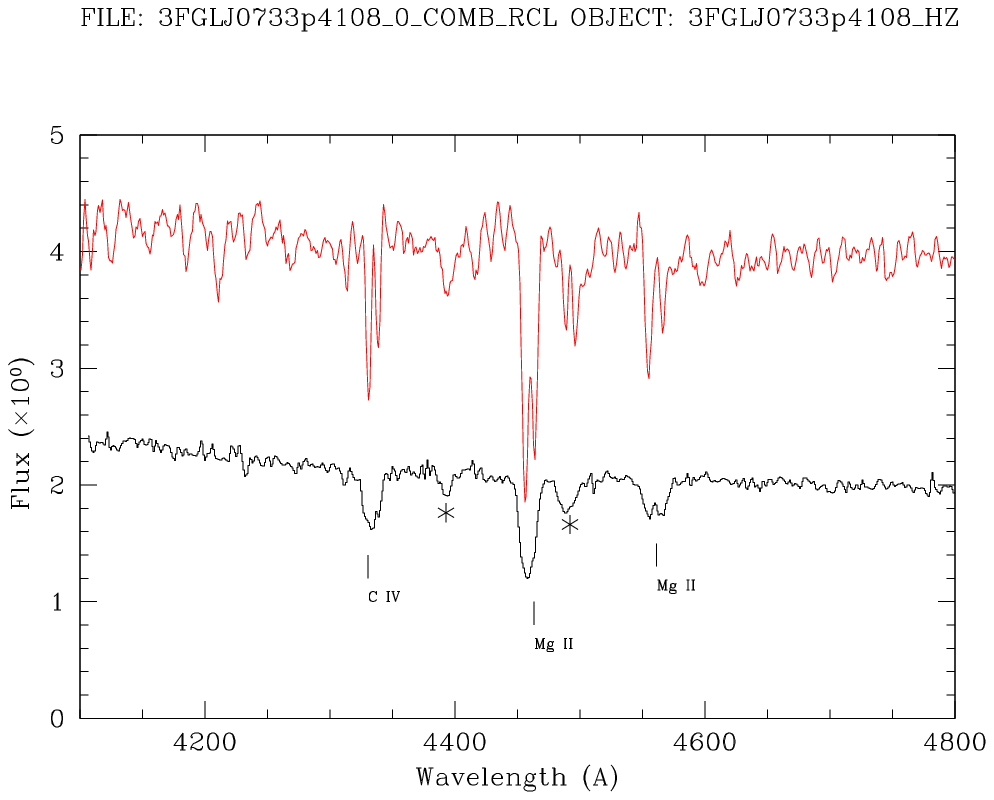}
\caption{The GTC optical spectrum of 4FGL~J0733.7+4110 (black line) in the blue region compared with that obtained by SDSS. The three main absorption features in the range 4200-4600 ~\AA~ are clearly doublets that support the identification as CIV 1550 and MgII 2800 intervening systems (see text for details).} 
\label{fig_0733}
\end{figure}

\item[] \textbf{4FGL~J0753.0+5353}:  
In the optical spectrum reported by \citet{shaw2013}, no significant emission and/or absorption features were detected, and a lower limit of z~$>$~0.42 is proposed.
In our GTC spectrum, we clearly detect a broad emission line at 5394~\AA, which, if attributed to MgII~2800, corresponds to a redshift z~=~0.927. 
Additionally, we identify an intervening MgII absorption doublet at 5176~\AA, associated with a system at z~=~0.851.
\\

\item[] \textbf{4FGL~J0754.4-1148}:   
Based on a featureless optical spectrum, \citet{shaw2013} proposed a lower limit to z~$>$~0.37 for the source. 
Our optical spectrum shows a continuum well described by a power law shape and it is featureless. No lines are detected down to EW~$>$~0.5~\AA.
\\

\item[] \textbf{4FGL~J0800.9+4401}:  
The spectrum presented by \citet{shaw2013} shows no detectable spectral features, from which a lower limit of z~$>$~0.51 was derived.
Two independent spectra, obtained by the SDSS, report tentative redshift estimates of z~=~1.07 (DR7) and z~=~1.38 (DR16). However, neither displays convincing spectral features to support these values. 
In our GTC spectrum, a narrow emission line is detected at 5852~\AA. The most plausible identification is with [OII]~3727~\AA, corresponding to a redshift z~=~0.57.
 An alternative identification of this feature as [O~III] $\lambda$5007 is ruled out by the signal-to-noise ratio: the primary line is sufficiently intense that the secondary [O~III] component at $\lambda$4959 would have been clearly detected.
This redshift  needs further confirmation from other lines.
\\

\item[] \textbf{ 4FGLJ0801.1+1335 }: 
An optical spectrum provided by  SDSS is noisy and featureless, exhibiting a power law continuum. The BZCAT catalogue lists a redshift of z~=~1.042.
Our GTC spectrum reveals an absorption doublet at 5360~\AA, identified as an MgII~2800 intervening system at z~=~0.916. No emission lines with EW~$>$~2~\AA~ are found, setting a lower limit on the redshift of the source of z~$>$~0.916.
\\

\item[] \textbf{4FGLJ0823.3+2224}:  
A redshift z~=~0.951 was previously proposed based on a single emission line at $\sim$5460~\AA~ interpreted as MgII~2800 in a low S/N spectrum \citep{stickel1993}. 
In our high quality (S/N$\sim$140) optical spectrum, the emission feature is not confirmed, and no lines with EW~$>$~1~\AA~ are detected across the observed range. The redshift of the source therefore remains undetermined.
\\

\item[] \textbf{5BZB~J0839+0319}:  
 This target (also known as PKS0837+035 ) was observed by \citet{drinkwater1997}. An uncertain redshift z=1.57 is reported, with no spectral features identified.
Our GTC spectrum is characterized by a power law continuum and shows a clear absorption doublet at 5220~\AA~, identified as intervening MgII~2800 at z~=~0.864. 
A weaker absorption feature is also detected at 4440~\AA~, which we attribute to FeII at the same redshift. These absorptions set a lower limit on the source redshift of z~$>$~0.864.
No clear emission lines are detected at the previously proposed tentative redshift z~=~1.57.
\\

\item[] \textbf{4FGL~J0849.5+0456}: 
The SDSS automatic analysis of the optical spectrum provides tentative redshift estimates of z~=~1.07 (DR7) and z~=~1.22 (DR16), but no convincing spectral features are evident. 
In our GTC spectrum, we detect a narrow emission line at 6500~\AA~ (EW~=~0.7~\AA~) that if interpreted as [OII]~3727~\AA~, corresponds to z~=~0.744.  
 An alternative identification of this feature could be as [O~III] $\lambda$5007. In this case we would expect to  detect also the secondary [O~III] component at $\lambda$4959. 
\\

\item[] \textbf{4FGL~J0902.4+2051}: 
A tentative redshift of z~=~1.52 has been proposed from the SDSS spectrum, although no convincing emission/absorption lines are present.
Our GTC spectrum displays a smooth continuum consistent with a power-law shape ($\alpha$$\sim$-0.7) and reveals a clear absorption doublet at 4770~\AA~.
We interpret this feature as the intervening MgII~2800~\AA~ doublet at z~=~0.706, which sets a lower limit to the source redshift of z~$>$~0.706.
\\

\item[] \textbf{4FGL~J0921.7+2336}: 
The SDSS spectrum appears featureless, though a highly uncertain redshift of z~=~1.38 was suggested.
Our higher quality (S/N$\sim$200) GTC data confirm that the optical spectrum remains featureless across the 3770–7750~\AA~ range, with the exception of two faint absorption features (EW$\sim$0.5~\AA~) at 6006 and 6060~\AA~. These lines are consistent with the CaII doublet and may originate from the host galaxy at z~=~0.527, or alternatively from an intervening absorber. However, no nearby galaxies are evident in the field that could be responsible for the absorption.
 This redshift  needs further confirmation from other lines.
\\

\item[] \textbf{4FGLJ0926.4+5412}: 
A tentative redshift of z~=~0.852 was reported from the SDSS spectrum.
The GTC optical spectrum (S/N$\sim$50) shows a featureless power law continuum ($\alpha$$\sim$--1) with no lines detected above EW~$>$~1~\AA.
\\

\item[] \textbf{4FGL~J1007.0+3455}: 
The BZCAT catalogue lists a tentative redshift of z~=~0.612.
The only available optical spectrum, in literature from SDSS, is featureless with a continuum following a power law shape. Our higher quality GTC spectrum (S/N$\sim$70) confirms the absence of any detectable features with EW~$>$~0.7~\AA. The continuum is well fitted by a power law with spectral index $\alpha$~=~--1.3. 
\\

\item[] \textbf{4FGL~J1054.5+2211}: 
A featureless optical spectrum reported by \citet{shaw2013}  (g=17.5) led to a lower redshift limit of z~$>$~0.51. Two additional spectra obtained by SDSS (DR7 and DR16) are also featureless. 
 The object was also observed by \cite{Alvarez-Crespo2025} in September 2023 (R=17.2), and no lines were detected. Our GTC observations (g=17.6) confirm the absence in the optical spectrum of any spectral lines (EW~$>$~0.3~\AA), with the continuum well described by a power law shape. 
 It is worth noting that the available SDSS spectra exhibit a significant flux variability. 
\\

\item[] \textbf{4FGLJ1110.5-1836}:
A redshift of z~=~1.56 was estimated by \citet{rajagopal2020} using a 
photometric technique based on six \textit{Swift}/UVOT filters and the SDSS \textit{griz} optical filters on the SARA–CTIO telescope. 
A featureless optical spectrum was obtained by \citet{shaw2013}, suggesting a lower limit z~$>$~0.5. Our GTC data confirms the featureless nature of the optical spectrum, with EW~$<$~0.8~\AA~ for the lines across the observed range. The continuum is well described by a power law ($\alpha$$\sim$--1).
\\

\item[] \textbf{4FGLJ1132.7+0034}: 
An uncertain redshift of z~=~1.14 was suggested from the SDSS-DR16 spectrum. In another optical spectrum obtained by \citet{shaw2013}, a narrow emission line at $\sim$6250~\AA~ was identified as [OII] at z~=~0.678, with a fainter emission feature at 8400~\AA~ attributed to [OIII] at the same redshift. Our GTC spectrum is dominated by a power law continuum. We confirm the [OII] emission at 6252~\AA~ with EW~=~0.8~\AA~, while the [OIII] line falls outside our observed spectral range. Additionally, weak CaII absorption lines from the host galaxy are detected.
Thus we confirm the redshift is z~=~0.678. 
\\

\item[] \textbf{4FGL~J1151.5-1347}:
The spectrum reported by \citet{shaw2013} shows many narrow absorption features superimposed on a power law continuum. A lower redshift limit of z~$>$~0.838 was suggested based on an intervening MgII system at $\sim$5150~\AA. Our GTC spectrum confirms the presence of this MgII absorber at z~=~0.838. At the same redshift we detect FeII at 4380 and 4779 ~\AA.~ In addition, other absorption features are detected at 4310, 4755, 4958 and 5245~\AA~, although their identification remains unknown.
\\

\item[] \textbf{4FGL~J1155.8+6137}: 
The 4LAC catalogue lists a redshift of z~=~1.5.
The SDSS optical spectrum shows a featureless continuum well described by a power law.
Our GTC observations confirm the absence of detectable lines, with the continuum remaining featureless and following a power law shape.
No emission or absorption features are detected with EW~$>$~0.5~\AA.
\\

\item[] \textbf{4FGL~J1215.1+5002}: 
Although the optical spectrum presented by \citet{shaw2013} is featureless, it provides a lower limit on the redshift of z~$>$~0.54.
Two further SDSS spectra also appear without significant lines.
Our high-quality GTC spectrum (S/N$\sim$160) reveals a featureless continuum well described by a power law.
No emission or absorption lines with EW~$>$~0.5~\AA~ are detected across the observed wavelength range.
\\

\item[] \textbf{4FGL~J1302.8+5748}: 

 A redshift z~=~1.088 is given by \cite{healey2008} but no spectra are shown.
In our GTC optical spectrum, we detect a weak broad emission line at 5844~\AA~ (EW$\sim$7~\AA~), accompanied by another broad emission at 3962~\AA~ and a fainter feature at 4850~\AA~. 
The three emission lines are identified as MgII, CIII] and CII] respectively, consistent with a redshift of z~=~1.088. 
\\

\item[] \textbf{4FGL~J1303.0+2434}:
A redshift z~=~0.993 is reported in \citet{glikman2007}, although no reference to the spectrum is provided.
In our GTC spectrum, we clearly detect a broad emission line (EW~=~2.5~\AA~) at 5572~\AA, which we identify as MgII~2800~\AA~, confirming the redshift of z~=~0.993. 
An absorption doublet is also detected at 4945~\AA~, attributed to an intervening MgII~2800 system at z~=~0.768. 
Two additional absorption lines (EW~=~0.3~\AA~) of the same system are detected at 4213 and 4599~\AA~ from FeII~2383 and FeII~2600, respectively. The spectral shape is well described by a power law with spectral index $\alpha$$\sim$-0.9.
\\

\item[] \textbf{4FGLJ1312.4-2156 }: 
The optical spectrum obtained by \citet{blades1980} revealed two intervening absorption systems at z~=~1.1361 and z~=~1.489, both attributed to CIV~1550.
\citet{shaw2013} later suggested a tentative redshift z~=~1.6 based on a very weak feature at $\sim$7280~\AA~ attributed to Mg~II~2800. 
In our high-quality spectrum (S/N$\sim$190), we do not confirm this line, which would in any case be strongly affected by the telluric absorption band near 7200~\AA~. No emission features are detected above EW~$>$~0.3~\AA~.
We clearly detect a narrow, strong absorption line at 3855~\AA~ (EW~$\sim$~10~\AA~), confirming the presence of the intervening system at z~=~1.49.
The continuum follows a power-law shape with a spectral index of $\alpha$$\sim$-0.4.
\\

\item[] \textbf{5BZB~J1423+1412}: 
A redshift z~=~0.768 was proposed  for SDSSJ142330.7+141247  from a SDSS noisy spectrum, although no clear emission features are visible. Our GTC optical spectrum does not confirm the value. We detect a broad emission line at 7109~\AA~, but the detection remains somewhat uncertain because it is partially contaminated by the telluric absorption.
If this feature corresponds to MgII~2800, the redshift is z~=~1.53. 
In addition, a distinct clear absorption line is also found at 4203~\AA~, which is most likely due to an intervening MgII~2800 at z~=~0.501. 
\\

\item[] \textbf{5BZB~J1437+4717}: 
From the very noisy SDSS spectrum  of SDSSJ143716.1+471726, a redshift of z~=~0.943 was suggested.
Our GTC spectrum is dominated by a power law continuum ($\alpha$$\sim$-1.1), with no significant spectral features detected with EW~$>$~1.5~\AA. 
The redshift of the source therefore remains unknown. 
\\

\item[] \textbf{4FGL~J1438.6+1205}: 
The SDSS automatic analysis of the optical spectrum suggests a possible weak emission at about 4200~\AA. This feature yields a redshift of z~=~0.847, if identified as CII].
We do not confirm this feature in our GTC spectrum therefore, the redshift of the source is unknown.
\\

\item[] \textbf{4FGL~J1500.7+4752}: 
A tentative redshift z~=~1.059 is reported in a note of the NED, based on an unpublished paper by Vermeulent et al. (1996), where two emission lines identified as MgII and [OII]. 
In our GTC spectrum, we detect only a weak emission line at 5765~\AA~ (EW$\sim$~3~\AA~), which, if attributed to MgII~2800, confirms the proposed redshift of z~=~1.059. The continuum is rather flat, with a spectral index of $\alpha$~=~ --0.2.
\\

\item[] \textbf{4FGL~J1503.3+1651}:
This source was observed at two different epochs by the SDSS. 
From these spectra, uncertain redshifts of z~=~0.822 (DR7) and z~=~0.678 (DR16)  were suggested, although no significant emission or absorption lines are clearly visible 
 at the proposed redshift. Note also that the GTC spectrum was obtained at a flux level that is a factor $\sim$ 1.7 higher than that obtained by SDSS.
Our GTC optical spectrum is characterized by a linear continuum without detectable emission lines with EW~$>$~0.5~\AA~. 
Conversely, several intervening absorption systems are detected at 4694, 5122, and 5508~\AA~, corresponding to FeII~2383, FeII~2600 and MgII~2800 at z~=~0.970.
Additional three intervening absorption features are present at 4620, 4680, and 5097~\AA with EW~=~1.1, 0.6, and 0.7~\AA~, respectively; the latter two are consistent with FeII at z~=~0.970. From careful inspection of the DR 16 SDSS spectrum these absorption doublets are also visible. 
The absorption systems set for the target a lower redshift limit of z~$>$~0.970.
\\

\item[] \textbf{4FGL~J1509.7+5556}: 
The optical spectra obtained by \citet{shaw2013} and by the SDSS are both featureless, with \citet{shaw2013} proposing a redshift lower limit of z~$>$~0.5. 
Our high quality GTC spectrum (S/N$\sim$200) likewise shows no detectable spectral features down to EW~$>$~0.2~\AA~ across the entire observed wavelength range. The continuum is featureless and well represented by a power law with spectral index $\alpha$~=~-1.2.
\\

\item[] \textbf{4FGL~J1529.2+3812}: 
Both SDSS spectra appear featureless, and no reliable redshift determination is available. 
Our GTC spectrum displays a linear continuum described by a power law shape ($\alpha$$\sim$-1.1) without any evident emission or absorption lines.    
Consequently, the redshift of the source remains unknown.
\\

\item[] \textbf{4FGL~J1530.9+5736}: 
Two SDSS optical spectra are available, but neither shows identifiable spectral features. 
Our GTC spectrum reveals a continuum well described by a power law ($\alpha$$\sim$--0.7) and no detectable emission lines with EW~$>$~0.5~\AA. 
However, we clearly detect an absorption system that includes MgII~2800 and FeII~2600, corresponding to at z~=~1.0727. 
This establishes a lower limit to the redshift of this target at z~$>$~1.073.
\\

\item[] \textbf{4FGL~J1549.6+1710}: 
A low quality optical spectrum was obtained by the SDSS and no convincing emission or absorption lines are found. Our GTC spectrum (S/N$\sim$80) is characterized by a featureless continuum, well described by a power law shape with spectral index of $\alpha$$\sim$-0.6. 
No spectral features with EW limit $>$~0.5~\AA~ are detected. 
\\

\item[] \textbf{5BZB~J1555+2141}: 
Two SDSS spectra obtained for SDSSJ155500.5+214159  suggest tentative redshifts of z~=~0.8645 (DR7) and z~=~0.802 (DR16). 
Our GTC spectrum exhibits a power law continuum with $\alpha$$\sim$-0.9.
A weak emission line is detected at 6947~\AA~, although it is partially contaminated by the telluric absorption band. 
This feature is also marginally visible in the SDSS spectra.
If interpreted as [OII]~3727~\AA~, it corresponds to a redshift of z~=~0.864.
 In Figure \ref{fig:zoom} we show this feature after correction of the telluric absorption.
\\

\item[] \textbf{4FGL~J1559.9+2319}: 
From a noisy SDSS spectrum, an uncertain redshift z~=~1.034 was suggested but no clear emission/absorption features are visible.
In our very good spectrum (S/N$\sim$~250), we detect an intervening MgII~2800 absorption doublet at 4344~\AA, establishing a lower limit on the redshift of z~$>$~0.554. No emission lines are observed down to EW~$>$~0.3~\AA.
\\

\item[] \textbf{4FGL~J1630.7+5221}: 
\citet{shaw2013} reported a featureless spectrum, setting a lower redshift limit of z~$>$~0.43. 
We confirm the absence of any spectral features in out GTC optical spectrum with a limit of any absorption or emission line at EW~$>$~0.3 ~\AA.
\\

\item[] \textbf{4FGL~J2145.5+1006}: 
A tentative redshift z~=~0.665 is given from the noisy SDSS-DR16 spectrum, although no clear emission/absorption features are correctly identified. 
Our GTC spectrum is characterized by a power law continuum with distinct absorption systems at z~=~0.661, attributed to intervening MgII~2800~\AA~ and FeII~2600~\AA~ material.
Another absorption feature is detected at 4297~\AA~, located near the FeII~2600~\AA~ line. 
These features indicate that the redshift of the source is z~$>$~0.661.
\\

\item[] \textbf{4FGL~J2152.5+1737}: 
The spectrum presented by \citet{shaw2013} is well described by a power law continuum.
In that noisy spectrum, a possible weak emission is found at $\sim$5247~\AA~  and identified as MgII~2800~\AA~ yielding z~=~0.874. 
In our GTC spectrum, the MgII~2800 line is not detected. 
However, a narrow emission is tentatively observed at 6970~\AA~. 
If identified as [OII]~3727, this line implies a redshift of z~=~0.870.
\\

\item[] \textbf{4FGL~J2206.8-0032}: 
A spectrum obtained by \citet{shaw2013} (g=19.1)  exhibits a power law continuum with an emission feature at $\sim$5750~\AA~  that was identified as MgII~2800 at z~=~1.053. In our GTC spectrum (g=19.3), we do not confirm this emission line. At its expected wavelength, we set an upper limit of EW~$<$~0.3. Thus, our data indicate a featureless spectrum. 
\\
\item[] \textbf{5BZB~J2339+0534}: 
A tentative redshift of z~=~0.74 is provided for MS2336.5+0517 by \citet{rector2000}.
Our GTC optical spectrum exhibits a power law continuum and it is featureless with a minimum EW~=~1.0~\AA~.
This object  was resolved in an optical image obtained by \citet{falomo_kotilainen1999}. From image decomposition the host galaxy has a magnitude r=18.6 and an effective radius Re=0.6 arcsec. Assuming the host galaxy is a typical BL Lac host galaxy (see e.g. \citet{sbarufatti2005b} ) the redshift of this source would be in the range 0.4 to 0.6.
\\
\end{itemize}

\section{Extra targets}
\label{sec:extra}
We are reporting here spectroscopy of three additional BLL that do not satisfy the selection criteria but were obtained during the same observation run because of particular interest.
\\

\subsection{5BZBJ1107+5010} 
\begin{figure}[h]
\begin{center}
\includegraphics[width=9cm]{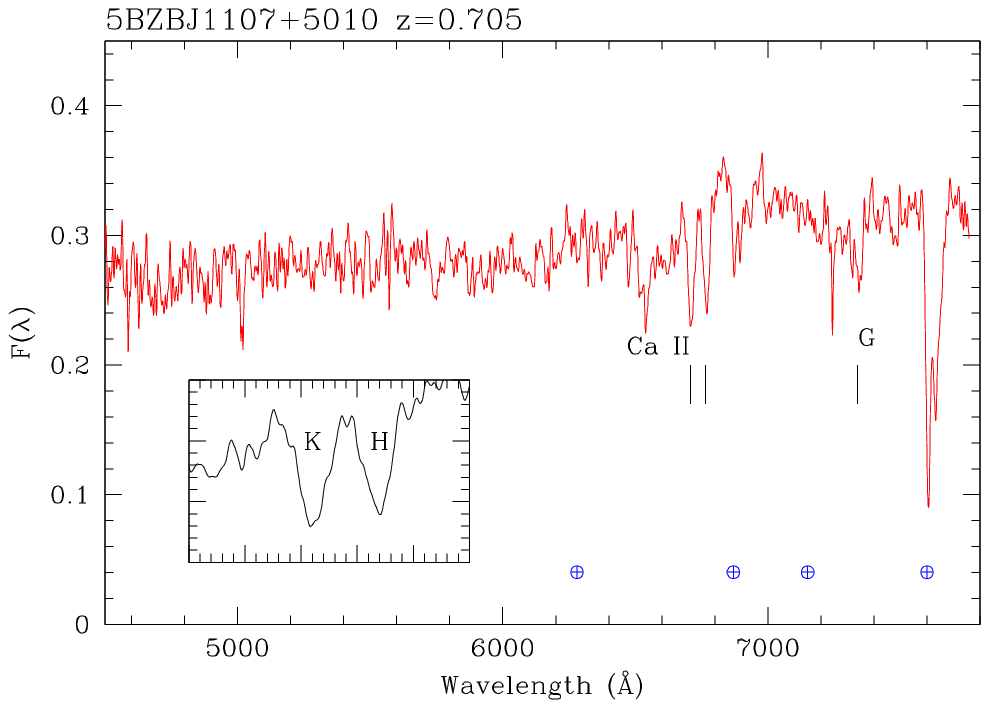  }
\caption{GTC optical spectrum of the BLL 5BZBJ1107+5010.
A strong absorption doublet of CaII and G band are detected at z=0.705. In the inset box a zoom of the CaII doublet is shown. }
\label{fig_1107}
\end{center}
\end{figure}

\textit{}
This target  ( SDSSJ110704.8+501038)   appears as a BLL in the BZCAT, however, no detection at  high energy (gamma and X-ray) was found. In a noisy spectrum obtained by SDSS a redshift z=0.705 was proposed superposed onto a rather flat optical continuum. We obtained an optical spectrum that confirms this redshift based mainly on the clear detection of the CaII H and K absorption doublet at 6710, 6770 ~\AA and G band. These two lines are particularly strong with EW~=~7.8 and 6.8~\AA for K and H, respectively (see Figure \ref{fig_1107} ). To explain the strength of these lines the  host galaxy would be rather luminous ( M(R)$\sim$~-23.2 ) and the nucleus/host flux ratio $\sim$~0.5 at wavelength close to CaII absorption lines. Moreover, the slope of the spectral slope would be rather steep ($\alpha$$\sim$~-3). 
\\

\subsection{4FGL~J1824.1+5651}
\begin{figure}[h]
\begin{center}
\includegraphics[width=9cm]{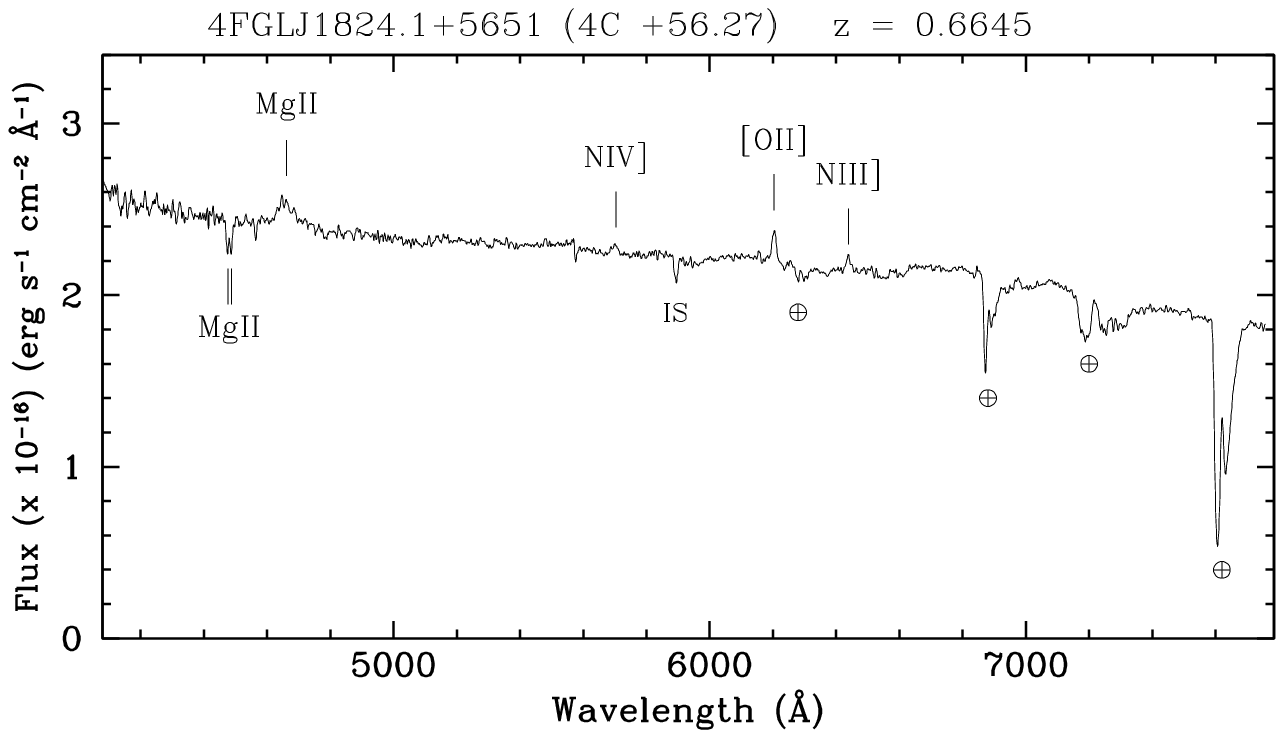}
\caption{The GTC optical spectrum of  4FGL~J1824.1+5651 (4C +56.27). Broad and narrow emission lines at detected at z~=~0.6645. Also an intervening absorption system of MgII~2800 is observed at z~=~0.601. }
\label{fig_1824}
\end{center}
\end{figure}

This source is a historically known blazar \citep[see e.g.][]{healey2007}. An initial redshift of $z = 0.663$ was proposed by \citet{lawrence1996} and later confirmed by \citet{stickel1993} through the detection of weak broad Mg\,\textsc{ii} and narrow [O\,\textsc{ii}]~$\lambda$3727~\AA\ emission lines. Additional optical spectra obtained by \citet{shaw2012} and \citet{torrealba2012} confirmed the presence of these lines and the corresponding redshift.  
The source exhibited a significant increase in gamma-ray activity in November 2024 \citep[see][]{giroletti2024ATel16913}. We obtained spectroscopic observations with GTC \citep[see][]{paiano2024ATel16925}, clearly detecting the broad Mg\,\textsc{ii}~$\lambda$2800~\AA\ line (EW = 3.5~\AA) and the narrow [O\,\textsc{ii}]~$\lambda$3727~\AA\ line. Two additional emission lines are identified at 5703~\AA\ and 6439~\AA, corresponding to [Ne\,\textsc{v}]~$\lambda$3426 and [Ne\,\textsc{iii}]~$\lambda$3869, respectively. From these features, we determine a firm redshift z~=~0.6645. We also find an intervening absorption system due to MgII~2800 at z~=~0.601 (see Figure \ref{fig_1824}).

\subsection{4FGL2134.2-0154}
\begin{figure}[H]
\begin{center}
\includegraphics[width=9cm]{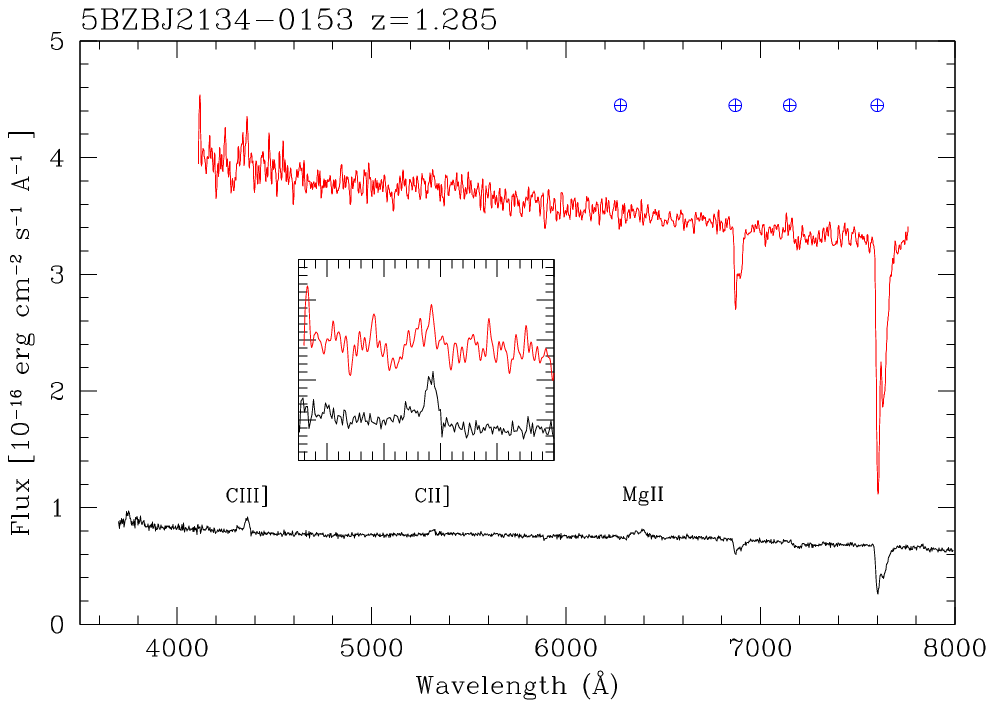}
\caption{ The GTC optical spectrum (upper red spectrum) of 5BZBJ2134-0153  obtained in 2018 compared with thatobtained in 2004 by \citet{sbarufatti2006} (lower black spectrum).  An increase of a factor 5 in flux is detected while the line fluxes are reduced by a factor $>$~4. The inset box shows a zoom of the CIII] emission line in the two spectra.}
\label{fig2134}
\end{center}
\end{figure}

An optical spectrum obtained by \citet{sbarufatti2006} confirmed the redshift z~=~1.285 proposed by \citet{rector2001} and \citet{drinkwater1997}. 
The redshift was based mainly on emission lines of CIII]~1909 and MgII~2800.
We observe this target in 2018 October since it was about a factor $\sim$~5 brighter than at the observation of 2004 by \citet{sbarufatti2006}. 
Our GTC spectrum is characterized by a power law continuum. No clear absorption or emission lines are apparent with EW greater than 2.5~\AA.  
It is of interest to compare this spectrum, with that obtained by \citet{sbarufatti2006} (see Fig.~\ref{fig2134}). 
In the latter spectrum, three emission lines are visibile yielding z~=~1.283. The strongest emission line is that of CIII]~1909 at $\sim$~4360~\AA  (EW$\sim$4.5~\AA). In the GTC spectrum obtained in the higher state, this feature is barely visible and the EW is less than 2~\AA (see Fig.~\ref{fig2134}).
On the other hand, the continuum shape is very similar (power law spectral index  is $\alpha$~=~-0.15 ( in 2018) and $\alpha$~=~-0.20 (in 2004).

\section{Tables}
\begin{table*}
 \begin{center}
 \caption{The sample of high z candidate blazars. } 
 \begin{tabular}{llllll} 
 \hline 
Object & RA & DEC &Date & E(B-V) & Ref(z)  \\  
 \hline 
4FGLJ0038.1+0012 & 00:38:08.48 & 00:13:36.5 & 2017-12-28 &   0.02  &         1,2  \\ 
4FGLJ0047.0+5657 & 00:47:00.40 & 56:57:42.0 & 2022-09-20 &    0.3  &         3  \\ 
4FGLJ0050.7-0929 & 00:50:41.3 & -09:29:05.0 & 2022-09-20 &   0.03  &         3  \\ 
4FGLJ0124.8-0625 & 01:24:50.48 & -06:25:00.9 & 2018-12-05 &   0.04  &         3  \\ 
4FGLJ0148.6+0127 & 01:48:33.8 & 01:29:01.0 & 2018-12-14 &   0.02  &         3  \\ 
4FGLJ0153.9+0823 & 01:54:02.77 & 08:23:51.1 & 2022-11-24 &   0.07  &         3  \\ 
4FGLJ0303.6+4716 & 03:03:35.24 & 47:16:16.3 & 2022-11-03 &   0.22  &         3  \\ 
5BZBJ0407+0742 & 04:07:29.09 & 07:42:07.5 & 2017-12-20 &   0.23  &          4  \\ 
4FGLJ0434.1-2014 & 04:34:07.91 & -20:15:17.1 & 2018-11-04 &   0.04  &         3  \\ 
4FGLJ0501.0+2424 & 05:01:06.86 & 24:23:16.4 & 2022-11-28 &   0.44  &         5  \\ 
5BZBJ0508+8432 & 05:08:42.36 & 84:32:04.5 & 2017-12-20 &   0.09  &          6,7  \\ 
4FGLJ0525.8-0052 & 05:25:54.63 & -00:51:40.4 & 2023-12-10 &   0.15  &      8  \\ 
4FGLJ0629.3-1959 & 06:29:23.76 & -19:59:19.7 & 2018-11-10 &   0.21  &         3  \\ 
4FGLJ0706.8+7742 & 07:06:51.33 & 77:41:37.0 & 2022-11-28 &   0.04  &         3  \\ 
5BZBJ0726+3734 & 07:26:59.52 & 37:34:23.0 & 2018-11-04 &   0.05  &          2,9  \\ 
4FGLJ0733.7+4110 & 07:33:46.8 & 41:11:20.0 & 2019-03-09 &   0.05  &         2,10  \\ 
4FGLJ0753.0+5353 & 07:53:01.38 & 53:52:59.6 & 2022-11-30 &   0.04  &         3  \\ 
4FGLJ0754.4-1148 & 07:54:26.4 & -11:47:16.9 & 2022-12-01 &   0.17  &         3  \\ 
4FGLJ0800.9+4401 & 08:01:08.28 & 44:01:10.2 & 2023-03-22 &   0.04  &         2,3  \\ 
5BZBJ0801+1336 & 08:01:15.05 & 13:36:42.6 & 2017-12-20 &   0.03  &          2,6  \\ 
5BZBJ0823+2223 & 08:23:24.76 & 22:23:03.4 & 2017-12-13 &   0.04  &          11  \\ 
5BZBJ0839+0319 & 08:39:49.25 & 03:19:54.2 & 2017-12-19/20 &   0.02  &          12  \\ 
4FGLJ0849.5+0456 & 08:49:32.55 & 04:55:07.9 & 2023-12-10 &   0.06  &         2  \\ 
4FGLJ0902.4+2051 & 09:02:26.91 & 20:50:46.4 & 2024-01-18 &   0.03  &         2  \\ 
4FGLJ0921.7+2336 & 09:21:45.38 & 23:35:48.2 & 2023-03-22 &   0.03  &         2  \\ 
5BZBJ0926+5411 & 09:26:38.8 & 54:11:26.5 & 2017-12-18 &   0.02  &          2  \\ 
4FGLJ1007.0+3455 & 10:06:56.47 & 34:54:45.1 & 2019-04-09 &   0.01  &         2,6  \\ 
4FGLJ1054.5+2211 & 10:54:30.56 & 22:10:55.5 & 2023-01-25 &   0.01  &          2,3  \\ 
5BZBJ1107+5010 & 11:07:04.8 & 50:10:38.0 & 2019-03-27 &   0.02  &          2  \\ 
4FGLJ1110.5-1836 & 11:10:27.69 & -18:35:51.2 & 2024-01-16 &   0.04  &          3,13  \\ 
5BZBJ1132+0034 & 11:32:45.61 & 00:34:27.8 & 2018-01-16 &   0.02  &          2  \\ 
4FGLJ1151.5-1347 & 11:51:30.00 & -13:47:51.0 & 2019-04-02 &   0.04  &         2  \\ 
4FGLJ1155.8+6137 & 11:55:48.40 & 61:35:54.0 & 2023-01-27 &   0.03  &         2,8  \\ 
4FGLJ1215.1+5002 & 12:15:00.80 & 50:02:16.0 & 2023-02-20 &   0.02  &         3, 8  \\ 
4FGLJ1302.8+5748 & 13:02:52.47 & 57:48:37.6 & 2023-03-27 &   0.01  &         14  \\ 
4FGLJ1303.0+2434 & 13:03:03.21 & 24:33:55.7 & 2018-03-16 &   0.02  &         15  \\ 
5BZBJ1312-2156 & 13:12:31.55 & -21:56:23.4 & 2023-03-24 &   0.10  &          3,16  \\ 
5BZBJ1423+1412 & 14:23:30.66 & 14:12:47.8 & 2018-03-26 &   0.02  &          2  \\ 
5BZBJ1437+4717 & 14:37:16.14 & 47:17:26.3 & 2018-03-26 &   0.02  &          2  \\ 
5BZBJ1438+1204 & 14:38:25.55 & 12:04:18.9 & 2018-04-29 &   0.02  &          2  \\ 
4FGLJ1500.7+4752 & 15:00:48.65 & 47:51:15.5 & 2023-03-28 &   0.02  &         14  \\ 
4FGLJ1503.3+1651 & 15:03:16.60 & 16:51:17.0 & 2020-05-11 &   0.04  &         2  \\ 
4FGLJ1509.7+5556 & 15:09:48.00 & 55:56:17.0 & 2023-07-19 &   0.01  &         3  \\ 
4FGLJ1529.2+3812 & 15:29:13.60 & 38:12:18.0 & 2023-07-21 &   0.01  &         2  \\ 
4FGLJ1530.9+5736 & 15:30:58.20 & 57:36:25.0 & 2023-02-20 &   0.01  &         2  \\ 
4FGLJ1549.6+1710 & 15:49:29.30 & 17:08:28.0 & 2020-05-10 &   0.03  &         2  \\ 
5BZBJ1555+2141 & 15:55:00.50 & 21:41:59.7 & 2018-04-29 &   0.05  &          2  \\ 
4FGLJ1559.9+2319 & 15:59:52.20 & 23:16:56.8 & 2023-04-25 &   0.05  &         2  \\ 
4FGLJ1630.7+5221 & 16:30:43.10 & 52:21:38.0 & 2022-10-01 &   0.02  &         3  \\ 
4FGLJ1824.1+5651 & 18:24:07.06 & 56:51:01.5 & 2018-08-07 &   0.05  &           18  \\ 
5BZBJ2134-0153 & 21:34:10.31 & -01:53:17.2 & 2018-10-18 &   0.05  &     19  \\ 
4FGLJ2145.5+1006 & 21:45:30.20 & 10:06:05.0 & 2022-10-03 &   0.06  &         2  \\ 
4FGLJ2152.5+1737 & 21:52:24.81 & 17:34:37.7 & 2018-08-30 &   0.09  &         3  \\ 
4FGLJ2206.8-0032 & 22:06:43.28 & -00:31:02.4 & 2018-11-26 &   0.10  &         3  \\ 
5BZBJ2339+0534 & 23:39:07.39 & 05:34:26.2 & 2018-09-29 &   0.07  &          17  \\ 
\hline 
\end{tabular} 
\label{tab:targets} 
\end{center} 
\footnotesize{\textbf{Note.} Column~1: Object name; Column~2: Right ascension; Column~3: Declination; Column~4: Date of observation; 
Column~5 : UT of the start of the observation ;Column~6: \textit{E(B-V)} from the NASA/IPAC Infrared Science Archive (\url{https://irsa.ipac.caltech.edu/applications/DUST/}); Column~7: Reference of the redshift reported in literature : 
 (1) \citet{sandrinelli2013}; (2) Sloan Digital Sky Survey (SDSS-DR7 or DR16); (3) \citet{shaw2013}; (4) \citet{sowards2003}; (5) \citet{sheng2024}; (6) Roma-BZCAT \citet{massaroe2015}; (7) \citet{stocke1997}; (8) 4LAC \citet{ajello2020}; (9) \citet{landoni2018}; (10) \citet{plotkin2010, plotkin2011}; (11) \citet{stickel1993}; (12) \citet{drinkwater1997}; (13) \citet{rajagopal2020}; 14) NASA/IPAC Extragalactic Database (NED); (15) \citet{glikman2007}; (16) \citet{blades1980}; (17) \citet{rector2000}; (18) \citet{lawrence1996}; (19) \citet{sbarufatti2006}. }
\end{table*}

\begin{table*}
 \begin{center}
 \caption{Results from GTC spectroscopy. } 
 \begin{tabular}{lllllll} 
 \hline 
Object & Filter & mag & S/N  & Redshift  & Note & Sp.Index   \\  
 \hline 
4FGLJ0038.1+0012 &     g &  20.21  &    15 &      *  &     B & -0.94   \\ 
4FGLJ0047.0+5657 &     g &  21.12  &    38 &      *  &     B & -2.33   \\ 
4FGLJ0050.7-0929 &     g &  16.08  &   377 &  0.636  &     A & -1.03   \\ 
4FGLJ0124.8-0625 &     g &  19.90  &    29 & >1.216  &     D & -0.50   \\ 
4FGLJ0148.6+0127 &     g &  19.80  &    29 &  0.944  &     A & -0.43   \\ 
4FGLJ0153.9+0823 &     g &  18.46  &   170 &  0.681  &     A & -0.79   \\ 
4FGLJ0303.6+4716 &     g &  17.30  &   248 &      *  &     B & -0.46   \\ 
 5BZBJ0407+0742 &     g &  20.82  &    15 &  1.141  &     C & -0.11   \\ 
4FGLJ0434.1-2014 &     g &  20.08  &    42 &  0.928  &     A & -0.65   \\ 
4FGLJ0501.0+2424 &     g &  20.09  &    29 &      *  &     B & -1.14   \\ 
 5BZBJ0508+8432 &     g &  19.05  &    21 & >1.340  &     A & -0.29   \\ 
4FGLJ0525.8-0052 &     r &  18.51  &    58 &      *  &     B & -0.56   \\ 
4FGLJ0629.3-1959 &     g &  20.25  &    29 &  1.730  &     A & -0.10   \\ 
4FGLJ0706.8+7742 &     g &  17.54  &   307 &      *  &     B & -0.82   \\ 
 5BZBJ0726+3734 &     g &  19.75  &    36 &  0.791  &     C & -0.98   \\ 
4FGLJ0733.7+4110 &     g &  18.46  &    99 &  >0.63  &     D & -1.08   \\ 
4FGLJ0753.0+5353 &     g &  19.67  &    71 &  0.927  &     C & -1.07   \\ 
4FGLJ0754.4-1148 &     g &  18.99  &    66 &      *  &     B & -1.16   \\ 
4FGLJ0800.9+4401 &     r &  19.65  &    45 &   0.57  &     C & -0.71   \\ 
 5BZBJ0801+1336 &     g &  19.61  &    63 & >0.916  &     D & -1.10   \\ 
 5BZBJ0823+2223 &     g &  17.83  &   140 &      *  &     B & -0.64   \\ 
 5BZBJ0839+0319 &     g &  20.73  &    18 &  >0.86  &     D & -1.74   \\ 
4FGLJ0849.5+0456 &     r &  18.25  &   118 &  0.744  &     C & -1.01   \\ 
4FGLJ0902.4+2051 &     r &  15.82  &   365 & >0.706  &     D & -0.66   \\ 
4FGLJ0921.7+2336 &     r &  17.93  &   200 &  0.527  &     C & -0.83   \\ 
 5BZBJ0926+5411 &     g &  19.31  &    50 &      *  &     B & -1.05   \\ 
4FGLJ1007.0+3455 &     g &  19.39  &    77 &      *  &     B & -1.27   \\ 
4FGLJ1054.5+2211 &     g &  17.56  &   247 &      *  &     B & -0.90   \\ 
4FGLJ1110.5-1836 &     r &  18.52  &    64 &      *  &     B & -1.02   \\ 
 5BZBJ1132+0034 &     g &  18.82  &    51 &  0.678  &     A & -0.73   \\ 
4FGLJ1151.5-1347 &     g &  20.38  &    25 & >0.838  &     A &  0.23   \\ 
4FGLJ1155.8+6137 &     g &  19.17  &    82 &      *  &     B & -0.84   \\ 
4FGLJ1215.1+5002 &     g &  18.46  &   162 &      *  &     B & -0.75   \\ 
4FGLJ1302.8+5748 &     r &  20.89  &    22 &  1.088  &     A & -0.22   \\ 
4FGLJ1303.0+2434 &     g &  18.59  &   154 &  0.993  &     A & -0.86   \\ 
 5BZBJ1312-2156 &     r &  16.90  &   189 &  >1.49  &     A & -0.42   \\ 
 5BZBJ1423+1412 &     g &  19.51  &    30 &   1.53  &     C & -1.13   \\ 
 5BZBJ1437+4717 &     g &  20.01  &    33 &      *  &     B & -1.07   \\ 
 5BZBJ1438+1204 &     g &  18.66  &    42 &      *  &     B & -1.51   \\ 
4FGLJ1500.7+4752 &     r &  19.60  &    34 &  1.059  &     A & -0.17   \\ 
4FGLJ1503.3+1651 &     g &  18.79  &    80 & >0.970  &     D & -1.26   \\ 
4FGLJ1509.7+5556 &     r &  17.85  &   201 &      *  &     B & -1.19   \\ 
4FGLJ1529.2+3812 &     r &  18.93  &    60 &      *  &     B & -1.09   \\ 
4FGLJ1530.9+5736 &     g &  19.45  &    69 & >1.073  &     A & -0.66   \\ 
4FGLJ1549.6+1710 &     g &  19.49  &    82 &      *  &     B & -0.64   \\ 
 5BZBJ1555+2141 &     g &  18.26  &    67 &  0.864  &     C & -0.86   \\ 
4FGLJ1559.9+2319 &     r &  17.34  &   250 & >0.553  &     D & -0.91   \\ 
4FGLJ1630.7+5221 &     g &  17.11  &   219 &      *  &     B & -0.59   \\ 
4FGLJ2145.5+1006 &     g &  19.60  &    56 & >0.661  &     D & -0.61   \\ 
4FGLJ2152.5+1737 &     g &  19.57  &    28 &  0.874  &     A & -0.24   \\ 
4FGLJ2206.8-0032 &     g &  19.35  &    50 &      *  &     B & -0.07   \\ 
 5BZBJ2339+0534 &     g &  20.09  &    25 &      *  &     B & -1.68   \\ 
\hline 
\end{tabular} 
\label{tab:results} 
\end{center} 
\footnotesize{\textbf{Note.} Column~1: Name of the target, Column~2: Observation filter, Column~3: Magnitude from our photometry, Column~4: Average Signal-to-Noise (S/N); 
Column~5: Seeing (arcsec); Column~6: Airmass; 
Column~: Total exposure time (sec);
Column~8: Redshift (*~=~featureless spectrum ); Column~9: Notes on the redshift: A~=~confirmed redshift or a lower limit; B~=~unknown redshift; C~=~new redshift; D~=~new lower limit of the redshift; Column~10: Spectral Index of the Power Law continuum.} 
\end{table*}

\begin{table*}
 \begin{center}
 \caption{Emission line measurements of BL Lacs. } 
 \begin{tabular}{llllll} 
 \hline 
Object name & $\lambda$  (\AA) & Ident & EW & Flux & Log(L)   \\  
 \hline 
 4FGLJ0050.7-0929 &   6095.5 &  [OII] &     0.1 &     1.8 &     41.49 \\ 
 4FGLJ0148.6+0127 &   7241.2 &  [OII] &     2.7 &     1.4 &     41.79 \\ 
 4FGLJ0153.9+0823 &   6271.6 &  [OII] &     0.6 &     1.1 &     41.36 \\ 
  5BZBJ0407+0742 &   5990.5 &   MgII &    17.0 &     9.5 &     42.84 \\ 
 4FGLJ0434.1-2014 &   7181.0 &  [OII] &     1.4 &     0.6 &     41.44 \\ 
 4FGLJ0434.1-2014 &   5395.0 &   MgII &     2.6 &     1.4 &     41.77 \\ 
 4FGLJ0629.3-1959 &   4245.0 &    CIV &     3.0 &     2.5 &     42.70 \\ 
 4FGLJ0753.0+5353 &   5396.3 &   MgII &     5.1 &     3.4 &     42.17 \\ 
 4FGLJ0800.9+4401 &   5850.4 &  [OII] &     2.8 &     1.3 &     41.25 \\ 
 4FGLJ0849.5+0456 &   6499.0 &  [OII] &     0.7 &     1.1 &     41.44 \\ 
  5BZBJ1132+0034 &   6252.4 &  [OII] &     0.7 &     0.9 &     41.26 \\ 
 4FGLJ1302.8+5748 &   5847.1 &   MgII &     6.6 &     0.9 &     41.74 \\ 
 4FGLJ1303.0+2434 &   5572.7 &   MgII &     2.7 &     4.5 &     42.36 \\ 
  5BZBJ1423+1412 &   7089.0 &   MgII &    10.0 &     5.0 &     42.87 \\ 
 4FGLJ1500.7+4752 &   5749.0 &   MgII &     2.5 &     1.2 &     41.86 \\ 
  5BZBJ1555+2141 &   6946.6 &  [OII] &     1.1 &     2.3 &     41.92 \\ 
 4FGLJ2152.5+17375 &   6967.7 &  [OII] &     2.9 &     2.5 &     41.97 \\ 
\hline 
\end{tabular} 
\label{tab:emlines} 
\end{center}  
\footnotesize{\textbf{Note.} Column~1: Name of the target; Column~2: Barycenter of the detected line; Column~3: Line identification; Column~4: Measured equivalent width (EW) in~\AA; Column-5: EW error in~\AA;
Column~6: Line flux ($\times$10$^{-16}$ erg cm$^{-2}$ s$^{-1}$); Column~7: Log of the line luminosity.\\}
\end{table*}

\begin{table*}
\begin{center}
\caption{Intervening absorption lines } 
\begin{tabular}{lllll} 
\hline 
Object name & $\lambda$ (\AA) & Line & EW   & z(abs) \\ 
\hline 
4FGLJ0124.8-0625 &   4679.4 &  MgII* &  3.9  & 0.672 \\
		 &   4730.9 &  MgII* &  3.9  & 0.690 \\
		 &   4832.0 &  MgII* &  2.9  & 0.726 \\
		 &   6196.8 &  MgII  &  1.9  & 1.216 \\
4FGLJ0509.6+8425 &   6541.8 &  MgII &  0.8  & 1.340 \\
                 &   6559.5 &  MgII &  0.7  & 1.340  \\
4FGLJ0629.3-1959 &   4220.7 &  MgII* & 4.3  & 0.507  \\
                 &   4554.7 &  MgII* & 1.9  & 0.627  \\
                 &   6389.3 &    ?  & 1.5  & ?  \\
                 &   6494.7 &    ?  & 2.7  & ?  \\
                 &   7049.5 &    ?  & 1.7  & ?  \\
                 &   7088.3 &    ?  & 2.5  & ?  \\
4FGLJ0733.7+4110 &   4333.4 &  CIV+  &  3.2  & 1.797 \\
                 &   4458.4 &  MgII+ &  5.7  & 0.594 \\
                 &   4560.6 & MgII+  &  3.1  & 0.629   \\
4FGLJ0753.0+5353 &   5175.9 &  MgII &  0.9  & 0.851  \\
                 &   5189.3 &  MgII &  0.7  & 0.851  \\
4FGLJ0801.1+1335 &   5358.5 &  MgII &  1.15 & 0.916  \\
                 &   5372.0 &  MgII &  1.13 & 0.916   \\
5BZBJ0839+0319   &   5210.5 &  MgII &  3.8  & 0.864   \\
                 &   5223.4 &  MgII &  4.5  & 0.864   \\
                 &   4438   &  FeII &  3.6  & 0.864  \\
4FGLJ0902.4+2051 &   4770.7 &  MgII &  0.3  & 0.706  \\
                 &   4783.0 &  MgII &  0.2  & 0.706  \\
4FGLJ1151.5-1347 &   5142.0 &  MgII & 5.0  & 0.839  \\
                 &   5154.5 &  MgII & 4.5  & 0.839  \\
4FGLJ1303.0+2434 &   4945.7 &  MgII & 1.1  & 0.769  \\
                 &   4958.4 &  MgII & 0.9  & 0.769   \\
                 &   4212   &  FeIIa & 0.6  & 0.769  \\
                 &   4599   &  FeIIb & 0.35  &  0.769  \\
5BZBJ1312-2156   &   3855.5 &  CIV+ & 10   &  1.49 \\
5BZBJ1423+1412   &   4203.3 &  MgII* & 5.7  & 0.501   \\
4FGLJ1503.3+1651 &   5510.0 &  MgII   & 1.8  & 0.970  \\
                 &   5524.3 &  MgII   & 1.4  & 0.970  \\
4FGLJ1530.9+5736 &   5795.7 &  MgII & 1.0  & 1.073  \\
                 &   5811.2 &  MgII & 0.8  & 1.073   \\
                 &   5392   &  FeIIb & 0.5  & 1.073  \\
4FGLJ1559.9+2319  &  4344.0 &  MgII  & 0.8  & 0.553  \\
                  &  4355.8 &  MgII  & 0.6  & 0.553  \\
4FGLJ2145.5+1006  &  4645.6 &  MgII & 3.2  & 0.661  \\
                  &  4657.0 &  MgII & 2.8  &  0.661   \\
                  &  4320   & FeIIb  & 2.1  & 0.661   \\
\hline 
\end{tabular} 
\label{tab:intervening} 
\end{center} 
\raggedright
\footnotesize{
MgII = MgII 2796,2803 doublet; FeIIa = FeII 2383 ;  
FeIIb = FeII 2600. \\
 (+) identification is confirmed by other literature  measurements of the doublet \\
(*) identification is not confirmed by the measurement of the doublet \\ 
Column~1: Name of the target; Column~2: Barycenter of the detected line; Column~3: Line identification (MgII indicates MgII 2796,2803 doublet; FeIIa indicates FeII 2383 and FeIIb is FeII 2600.); Column~4: Measured equivalent width (EW) in ~\AA~; Column~5: Redshift of the intervening system } 
\end{table*}

\begin{table*}
\label{tab:4lac}
 \begin{center}
 \caption{Sample of 57 BLL with reliable literature spectra } 
 \begin{tabular}{llrl} 
 \hline 
Name object &Other name &  Redshift  & Ref   \\  
 \hline 
4FGLJ0004.0+0840 &  SDSS       J000359.23+084138.1 & > 1.503 &  PS17b \\ 
4FGLJ0006.4+0135 &  NVSS            J000626+013611 &   0.787 &  PS17b \\ 
4FGLJ0021.6-0855 &  NVSS            J002142-090044 &   0.648 &   SDSS \\ 
4FGLJ0101.0-0059 &  NVSS            J010058-005547 & > 0.680 &   PR10 \\ 
4FGLJ0203.7+3042 &  NVSS            J020344+304238 &   0.761 &   SDSS \\ 
4FGLJ0209.9+7229 &    S5                  0205+722 &   0.895 &   VR96 \\ 
4FGLJ0238.6+1637 &   PKS                  0235+164 &   0.940 &   CR87 \\ 
4FGLJ0433.6+2905 &   MG2              J043337+2905 &   0.910 &   PS20 \\ 
 5BZBJ0755+3726 &  SDSS       J075523.11+372618.7 &   0.606 &   SDSS \\ 
4FGLJ0811.4+0146 &    OJ                       014 &   1.148 &   SB05 \\ 
4FGLJ0814.4+2941 &    RX              J0814.4+2941 &   0.703 &  PS17b \\ 
 5BZBJ0832+3300 &  SDSS       J083251.45+330010.8 &   0.672 &   SDSS \\ 
4FGLJ0836.9+5833 &  NVSS            J083705+583151 &   0.611 &   SDSS \\ 
 5BZBJ0855+2601 &  SDSS       J085506.39+260142.1 &   0.695 &   SDSS \\ 
 5BZBJ0920+3910 &  SDSS       J092015.67+391014.0 &   0.607 &   SDSS \\ 
 5BZBJ0924+5300 &  SDSS       J092405.83+530034.0 &   0.640 &   SDSS \\ 
 5BZBJ0930+3933 &  SDSS       J093056.84+393335.6 &   0.638 &   SDSS \\ 
 5BZBJ1000+5746 &  SDSS       J100050.22+574609.2 &   0.640 &   SDSS \\ 
4FGLJ1003.6+2605 &   PKS                   1000+26 &   0.929 &   SDSS \\ 
4FGLJ1012.3+0629 &  NRAO                       350 &   0.727 &   SB05 \\ 
4FGLJ1058.4+0133 &    4C                    +01.28 &   0.890 &   SDSS \\ 
4FGLJ1107.8+1501 &    RX              J1107.7+1502 & > 0.602 &   PR10 \\ 
4FGLJ1137.9-1708 &  NVSS            J113755-171031 &   0.600 &  PiS07 \\ 
4FGLJ1202.5+3852 &  NVSS            J120257+385147 & > 0.805 &   PR08 \\ 
4FGLJ1215.1+3513 &    7C                 1212+3524 &   0.883 &   SDSS \\ 
4FGLJ1224.9+4334 &    B3                  1222+438 &   0.957 &   SDSS \\ 
4FGLJ1250.6+0217 &   PKS                  1247+025 &   0.955 &   SA13 \\ 
4FGLJ1253.8+6242 &  1RXS          J125400.1+624303 & > 0.867 &   PR10 \\ 
4FGLJ1301.6+4056 &    RX              J1301.7+4056 & > 0.649 &   PR10 \\ 
4FGLJ1309.4+4305 &    B3                  1307+433 &   0.691 &   LM18 \\ 
4FGLJ1325.6-0227 &  1RXS          J132542.1-022800 & > 0.791 &   PR08 \\ 
4FGLJ1422.6+5801 &   1ES                  1421+582 &   0.635 &   BN98 \\ 
4FGLJ1427.0+2348 &   PKS                  1424+240 &   0.604 &  PS17a \\ 
4FGLJ1435.5+2021 &   TXS                  1433+205 &   0.748 &   SDSS \\ 
4FGLJ1442.0+4348 &  SDSS       J144207.15+434836.6 & > 0.673 &   PR10 \\ 
4FGLJ1450.8+5201 &  SDSS       J145059.99+520111.7 & > 2.470 &  PS17b \\ 
4FGLJ1451.4+6355 &    RX              J1451.4+6354 &   0.650 &   BN98 \\ 
4FGLJ1517.7+6525 &    1H                  1515+660 & > 0.702 &   BV99 \\ 
4FGLJ1540.7+1449 &    4C                    +14.60 &   0.605 &   SDSS \\ 
4FGLJ1546.5+1816 &   MG1              J154628+1817 & > 1.005 &   PR10 \\ 
4FGLJ1552.0+0850 &   TXS                  1549+089 & > 1.015 &   PR10 \\ 
4FGLJ1553.3+0600 &  NVSS            J155331+060143 &   0.619 &   SDSS \\ 
4FGLJ1643.0+3223 &  NVSS            J164301+322104 & > 1.028 &   PR08 \\ 
 5BZBJ1719+5524 &  SDSS       J171902.35+552434.5 &   0.626 &   SDSS \\ 
4FGLJ1747.9+4704 &    B3                  1746+470 &   1.484 &   VR96 \\ 
4FGLJ1800.6+7828 &    S5                  1803+784 &   0.680 &   PS20 \\ 
 5BZBJ1840+5452 &  87GB           183956.7+544928 &   0.646 &   LH11 \\ 
4FGLJ2156.9-0854 &  NVSS            J215650-085535 & > 1.017 &   PR08 \\ 
4FGLJ2225.5-1114 &   PKS                  2223-114 &   0.997 &   SB06 \\ 
4FGLJ2227.9+0036 &   PMN                J2227+0037 & > 1.093 &   PS21 \\ 
4FGLJ2236.2-1706 &   PKS                  2233-173 &   0.647 &   HI03 \\ 
4FGLJ2244.6+2502 &  NVSS            J224436+250345 &   0.650 &  PS17b \\ 
4FGLJ2244.9-0007 &  NVSS            J224448-000616 &   0.640 &   SA13 \\ 
4FGLJ2247.4-0001 &   PKS                  2244-002 &   0.949 &   SM13 \\ 
4FGLJ2255.2+2411 &   MG3              J225517+2409 & > 0.863 &   PS21 \\ 
4FGLJ2321.5-1619 &  NVSS            J232137-161935 &   0.694 &  PS17b \\ 
4FGLJ2357.4-0152 &   PKS                  2354-021 &   0.812 &   SB05 \\ 
\hline 
\end{tabular} 
\label{tab:4lac} 
\end{center}  
\footnotesize{\textbf{Note.} Column~1: Object name from 4LAC or BZCAT catalogues; Column~2: Other name, Column~3: Redshift; Column~4: Reference for the optical spectrum. References are listed according to the e following abbreviation: BN98: \citet{bade1998}, BV99: \citet{beckmann1999}, CR87: \citet{cohen1987}, HI03: \citet{hook2003}, LM18: \citet{landoni2018}, LH: \citet{landt2001}, PR08: \citet{plotkin2008}, PR10: \citet{plotkin2010}, PS17a: \citet{paiano2017tev}, PS17b: \citet{paiano2017high}, PS20: \citet{paiano2020fhl}, PS21: \citet{paiano2021sin1}, PiS07: \citet{piranomonte2007}, SA13: \citet{sandrinelli2013}, SB05: \citet{sbarufatti2005}, SDSS: (DR7 or DR16), SM13: \citet{shaw2013}, VR96: \citet{vermeulen1996} }
\end{table*}

\end{document}